\documentclass[twocolumn,showpacs,preprintnumbers,prd,floatfix]{revtex4} 
\usepackage{epsfig}
\usepackage{array}
\usepackage{graphicx}
\usepackage{dcolumn}% Align table columns on decimal point
\usepackage{bm}
\usepackage{epstopdf}
\usepackage{amssymb}
\usepackage{natbib}
\begin{document}
\title{Spin Evolution of Accreting Neutron Stars: Nonlinear Development of the R-mode Instability}
%\title{How do the nonlinear effects of the r-mode instability limit the maximum spin frequency of accreting neutron stars?}
% Limits on the spin frequency of accreting neutron star imposed by the nonlinear development of the r-mode instability
% How fast can accreting neutron stars spin?: limits imposed by the nonlinear development of the r-mode instability
%Spin Evolution of the Accreting neutron stars: Nonlinear development of the r-mode instability
%\title{Nonlinear Development of the r-mode Instability of Accreting Neutron Stars: Three Modes}
\author{Ruxandra Bondarescu, Saul A. Teukolsky, and Ira Wasserman}
\affiliation{Center for Radiophysics and Space Research, Cornell University, Ithaca, NY 14853}
%\date{}                                           % Activate to display a given date or no date
\begin{abstract}
The nonlinear saturation of the r-mode instability and its effects on the spin evolution of Low Mass X-ray Binaries (LMXBs) are modeled using the triplet of modes at the lowest parametric instability threshold. We solve numerically the coupled equations for the three mode amplitudes in conjunction with the spin and temperature evolution equations.  We observe that very quickly the mode amplitudes settle into quasi-stationary states that change slowly as the temperature and spin of the star evolve. Once these states are reached,  the mode amplitudes can be found algebraically and the system of equations is reduced from eight to two equations: spin and temperature evolution. The evolution of the neutron star angular velocity and temperature follow easily calculated trajectories along these sequences of quasi-stationary states. The outcome depends on whether or not the star will reach thermal equilibrium, where the viscous heating by the three modes is equal to the neutrino cooling ($H=C$ curve). If, when the  r-mode becomes unstable,  the star spins at a frequency below  the maximum of the $H=C$ curve,  then it will reach a state of thermal equilibrium. It can then either (1) undergo a cyclic evolution with a small cycle size with a frequency change of at most 10\%, (2) evolve toward a full equilibrium state in which the accretion torque balances the gravitational radiation emission, or (3) enter a thermogravitational runaway on a very long timescale of $\approx 10^6$ years.  If the star does not reach a state of thermal equilibrium, then a faster thermal runaway (timescale of $\approx 100$ years) occurs and the r-mode amplitude increases above the second parametric instability threshold. Following this evolution requires more inertial modes to be included. The sources of damping considered are shear viscosity, hyperon bulk viscosity and viscosity within the core-crust boundary layer. We vary proprieties of the star such as the hyperon superfluid transition temperature $T_c$, the fraction of the star that is above the threshold for direct URCA reactions, and slippage factor, and map the different scenarios we obtain to ranges of these parameters. We focus on $T_c \gtrsim 5 \times 10^9$ K where nonlinear effects are important. Wagoner \cite{Wagoner} has shown that a very low r-mode amplitude arises at smaller $T_c$. 
%when the evolutions are started on the part of the r-mode stability curve with positive slope. 
For all our bounded evolutions the r-mode amplitude remains small $\sim 10^{-5}$. The spin frequency of accreting neutron stars is limited by boundary layer viscosity to $\nu_{\rm{max}} \approx 800 Hz [S_{\rm{ns}}/(M_{1.4} R_6)]^{4/11} T_8^{-2/11}$.  Fast rotators are allowed for $[S_{\rm{ns}}/(M_{1.4} R_6)]^{4/11} T_8^{-2/11} \sim 1$ and we find that in this case the r-mode instability would be active for about 1 in 1000 LMXBs and that only the gravitational waves from LMXBs in the local group of galaxies could be detected by advanced LIGO interferometers.
\end{abstract}
\maketitle

\section{Introduction}
R-modes are oscillations in rotating fluids that are due to the Coriolis effect. 
They are subject to the classical Chandrashekar-Friedman-Shutz (CFS) instability \cite{C,FS}, which is driven by the gravitational radiation backreaction force. Andersson \cite{nils} and Friedman and Morsink \cite{Sharon} showed that, in the absence of fluid dissipation, r-modes are {\it linearly} unstable at all rotation rates. However, in real stars  there is a competition between internal viscous dissipation and gravitational driving \cite{LOM} that depends on the angular velocity $\Omega$ and temperature $T$ of the star. Above a critical curve in the $\Omega-T$ plane the $n=3, m=2$ mode, referred to as 'the r-mode' in this work, becomes unstable.  At first, an unstable r-mode grows exponentially, but soon it may enter a regime where other inertial modes that couple to the r-mode become excited and nonlinear effects become important. Roughly speaking, nonlinear effects first become significant as the amplitude passes its first parametric instability threshold, which is very low ($\sim 10^{-5}$). Modeling and understanding the nonlinear effects is crucial in determining (1) the final saturation amplitude of the r-mode and (2) the limiting spin frequency that neutron stars can achieve. The r-mode amplitude and the duration of the instability are among the main factors that determine whether the associated gravitational radiation could be detectable by laser interferometers on Earth.
%Previous work modeled the nonlinear effects unphysically by assuming a fixed r-mode amplitude.
% This phase curve, referred to as the r-mode stability curve through the rest of this paper, defines an angular velocity   as a function of temperature $\Omega(T)$ and stars spinning with angular velocities above this curve will have unstable r-modes.

The r-mode instability has been proposed as an explanation for the sub-breakup spin rates of both Low Mass X-ray Binaries (LMXBs)  \cite{Lars,NKS} and young, hot neutron stars  \cite{LOM,AKS}. 
The idea that gravitational radiation could balance accretion was proposed independently by Bildsten \cite{Lars} and Andersson {\it et al.} \cite{NKS}.  Cook, Shapiro and Teukolsky \cite{saul1, saul2} model the recycling of pulsars to millisecond periods via accretion from a Keplerian disk onto a bare neutron star with $M=1.4 M_\odot$ when $\Omega=0$. Depending on the equation of state they found that spin frequencies of between $\approx 670$ Hz and 1600 Hz could be achieved before mass shedding or radial instability set in (these calculations predated the realization that the r-mode instability could limit the spin frequency). For comparison, the highest observed spin rate of  millisecond pulsars is 716 Hz for PSR J1748-2446ad \cite{fsharp1,fsharp2}.  PSR B1937+21, which was discovered in 1982, was the previous fastest known radio pulsar with a spin rate of $642 \; \rm{Hz}$ \cite{dsharp}; that this ``speed" record stood for 24 years suggests that neutron stars rotating this fast are rare.  Moreover, based on a Bayesian statistical analysis of the spin frequencies of the 11 nuclear-powered millisecond pulsars whose spin periods are known from burst oscillations, Chakrabarty {\it et al.} \cite{deepto1}  claimed a cutoff limit of $\nu_{\rm{max}} = 760 \; \rm{Hz}$ (95\% confidence); A more recent analysis, which added two more pulsars to the sample, found $\nu_{\rm{max}} = 730\; \rm{Hz}$ \cite{deepto2}. 

At first sight, one might conclude that mass shedding or radial instability sets $\nu_{\rm{max}}$, and that it is just above the record $\nu = 716$ Hz determined for PSR J1748-2446ad.
However, the nuclear equations of state consistent with this picture all have rather large radii $\approx 16-17$ km for non-rotating 1.4 $M_\odot$ models; see Table 1 in Cook {\it et al.} \cite{saul1}. For these equations of state, the r-mode instability should lead to $\nu_{\rm{max}}$ somewhat below 716 Hz; see Eq.\ (\ref{freq}) in Sec.\ \ref{detection} below. Thus, the r-mode instability may prevent  recycling by accretion from reaching mass shedding or radial instability. In other words, the detection of the 716 Hz rotator is consistent with accretion spin-up mitigated by the r-mode instability only for equations of state for which mass shedding or radial instability would permit even faster rotation. Ultimately, this may be turned into useful constraints on nuclear equations of state. However, at present the uncertainty in the physics of internal dissipation is a significant hindrance in establishing such constraints.

Since a physical model to follow the nonlinear phase of the evolution was initially unavailable,
Owen {\it et al.} \cite{Owen1998} proposed a simple one-mode evolution model in which they assumed that nonlinear hydrodynamics effects saturate the r-mode amplitude  at some arbitrarily fixed value.
 According to their model, once this maximum allowed amplitude is achieved, the r-mode amplitude remains constant and the star spins down at this fixed amplitude (see Eqs. (3.16) and (3.17) in Ref.\ \cite{Owen1998}). They used this model to study the impact of the r-mode instability on the spin evolution of young hot neutron stars assuming normal matter. In their calculation they include the effects of shear viscosity and n-p-e bulk viscosity. They found that the star would cool to approximately $10^9$ K and spin down from a frequency close to the Kepler frequency to about $100 \; \rm{Hz}$ in a period of $\sim 1$ yr \cite{Owen1998}.  
%due to electron-electron scattering

Most subsequent investigations that did not perform direct hydrodynamic simulations used the one-amplitude model of Ref. \cite{Owen1998} for studying the r-mode instability.  Levin \cite{levin} used this model to study the limiting effects of the r-mode instability on the spin evolution of LMXBs, assuming an r-mode saturation amplitude of $\sim 1$; he adopted a modified shear viscosity to match the maximum LMXB spin frequency of $330\; \rm{Hz}$ known in 1999. Levin found that the neutron star followed a cyclic evolution in the $\Omega-T$ phase plane. 
The star spins up for several million years until it crosses the r-mode stability curve, whereupon the r-mode becomes unstable and the star is viscously heated for a fraction of a year until the r-mode reaches its saturation amplitude ($\sim 1$). At this point the spin and r-mode amplitude evolution equations are changed, following the prescription of Ref. \cite{Owen1998} to ensure constant amplitude. The star then spins down by emitting gravitational radiation for another fraction of a year until it crosses the r-mode stability curve again and the instability shuts off.  The time period during which the r-mode is unstable was found to be about $10^{-6}$ times shorter than the spin-up time, and Levin concluded that it is unlikely that any neutron stars in LMXBs in our galaxy are currently spinning down and emitting gravitational radiation.
 %and hence none will be detected by laser interferometers on earth. 
 However, following work by Arras {\it et al.} \cite{arras} showing that nonlinear effects become significant at small r-mode amplitude, Heyl \cite{Heyl} varied the saturation amplitude, and found that the duration of the spin-down depends sensitively on it.  %He used the total energy of the r-mode when nonlinear couplings become important as found by Arras et al. \cite{arras} to scale value for the saturation amplitude with the same one mode amplitude and spin evolution as Refs \cite{levin,LOM, Owen1998}. 
 He predicted that the unstable phase could be as much as 30\% of the cyclic evolution for an r-mode saturation amplitude of  $\alpha\approx 10^{-5}$, and that this would make some of the fastest spinning LMXBs in our galaxy detectable by interferometers on Earth.
% , whereas in the scenario of Ref. \cite{levin} the duty cycle was only of order $10^{-6}$. 
 % The fraction of the total cycle time the r-mode is unstable, also referred to as the duty cycle, is of order $10^{-6}$ for this evolution. 
 
 Jones \cite{Jones} and Lindblom and Owen \cite{LO} pointed out that if the star contains exotic particles such as hyperons (massive nucleons where an up or down quark is replaced with a strange quark), internal processes could lead to a very high coefficient of bulk viscosity in the cores of neutron stars. While this additional high viscosity coefficient could eliminate the instability altogether in newly born neutron stars \cite{Jones, LO, HLY, mohit}, Nayyar and Owen \cite{mohit} proposed that it would enhance the probability of detection of gravitational radiation from LMXBs by blocking the thermal runaway. 
 
 The cyclic evolution found by Levin \cite{levin} and generalized by Heyl \cite{Heyl} arises when shear or boundary layer 
viscosity dominates the r-mode dissipation. In the evolutionary picture of Nayyar and Owen \cite{mohit}, the r-mode 
first becomes unstable at a temperature where shear and boundary layer viscosity dominate, but the resulting thermal 
runaway halts once hyperon bulk viscosity becomes dominant. The key feature behind the runaway is that shear and 
boundary layer viscosities both decrease with increasing temperature, so the instability 
speeds up as the star grows hotter.  However, if the bulk viscosity is sufficiently large 
the star can cross the r-mode stability curve at a point where the viscosity is 
an increasing function of temperature. Such scenarios were studied by Wagoner \cite{Wagoner} for 
hyperon bulk viscosty with low hyperon superfluid transition temperature;
similar evolution was found for strange stars by Andersson, Jones and Kokkotas \cite{AJK}. 
In this picture, the star evolves near the r-mode stability curve 
 until an equilibrium between accretion spin-up and gravitational radiation spin-down is achieved. The value of the r-mode amplitude remains below the 
lowest instability threshold found by Brink {\it et al.} \cite{Jeandrew1, Jeandrew2, Jeandrew3} 
for modes with $n<30$, and hence in this regime nonlinear effects may not play a role.
        
 Schenk {\it  et al.} \cite{Schenk} developed a formalism to study the nonlinear interaction of the r-mode with other inertial modes. They assumed a small r-mode amplitude and treated the oscillations of the modes with weakly nonlinear perturbation theory via three-mode couplings.  This assumption was tested by Arras {\it et al.} \cite{arras} and Brink {\it et al.} \cite{Jeandrew1, Jeandrew2, Jeandrew3}. Arras {\it et al.} proposed that a turbulent cascade would develop in the strong driving regime. They estimated that r-mode amplitude was small and could have values between $10^{-1}-10^{-4}$.  Brink {\it et al.} modeled the star as incompressible and calculated the coupling coefficients analytically.  They computed the interaction of about 5000 modes via approximatively 1.3 million couplings of the $10^9$ possible couplings among the modes with $n \le 30$. The couplings were restricted to mode triplets with a fractional detuning $\delta \omega/(2 \Omega)  < 0.002$ since near-resonances promote modal excitation at very small amplitudes. Brink {\it  et al.}  showed that the nonlinear evolution saturates at a very small amplitude, generally comparable to the lowest parametric instability threshold that controls the initiation of energy sharing among the sea of inertial modes. However, Brink {\it et al.} did not model accretion spin-up or neutrino cooling in their calculation and only included minimal dissipation via shear viscosity.
  
In this paper we begin a more complete study of the saturation of the r-mode instability including accretion spin up and neutrino cooling.  We use a simple model in which we parameterize uncertain properties of the star such as the rate at which it cools via neutrino emission and
the rate at which the energy in inertial modes dissipates via boundary
layer effects \cite{BU} and bulk viscosity. In order to exhibit the variety of 
possible nonlinear behaviors, we explore a range of models with 
different neutrino cooling and viscous heating coefficients by 
varying the free parameters of our model. In particular, we vary:
 (1) the slippage factor $S_{\rm{ns}}$, which regulates the boundary layer viscosity, between 0 and 1  (see for example \cite{LU,YL,GA} for some models of the interaction between the oscillating fluid core and an elastic crust)  ; (2) the fraction of the star that is above the density threshold for direct URCA reactions $f_{\rm{dU}}$, which is taken to be between 0 (0\% of the star cools via direct URCA) and 1 (100\% of the star is subjected to direct URCA reactions), and in general depends on the equation of state used; and (3) the hyperon superfluidity temperature $T_c$, which is believed to be between $10^9-10^{10}$ K  (We use a single, effective $T_c$ rather than modelling its spatial
variation.) We focus on $T_c \gtrsim 5 \times 10^9$ K for which nonlinear effects are important. For low $T_c \lesssim 3 \times 10^9$ K, Wagoner \cite{Wagoner} showed that the evolution reaches a steady state at amplitudes below the lowest parametric instability threshold found by Brink {\it et al.} \cite{Jeandrew3}.  It is important to note that all our evolutions start on the part of the r-mode stability curve that decreases with temperature and that the bulk viscosity does not play a role in any of our bound evolutions. %We are able to provide a complete map of this parameter space to the different evolution scenarios obtained.

We include three modes: the r-mode at $n=3$ and the two inertial modes at  $n=13$ and $n=14$ that become unstable at the lowest parametric instability threshold found by Brink {\it et al.} \cite{Jeandrew3}.  We evolve the coupled equations for the three-mode system numerically in conjunction with the spin and temperature evolution equations. The lowest parametric instability threshold provides a physical cutoff for the r-mode amplitude. In all cases we investigate, the growth of the r-mode is initially halted by energy transfer to the two daughter modes. We observe that the mode amplitudes settle into a series of quasi-stationary states within a period of a few years after the spin frequency of the star has increased above the r-mode stability curve. 
These quasi-stationary states are algebraic solutions of the three-mode amplitude equations (see Eqs.\
 (\ref{stationarySol})) and change slowly as the spin and the temperature of the star evolve. Using these solutions for the mode amplitudes, one can reduce the eight evolution equations (six for the real and imaginary parts of the mode amplitudes, which are complex \cite{Schenk}; one for the spin, and one for the temperature) to two equations governing the rotational frequency and the temperature of the star. Our work can be regarded as a minimal physical model for modeling amplitude saturation realistically. 

The outcome of the evolution is crucially dependent on whether the star can reach a state of thermal equilibrium. This can be predicted by finding the curve where the viscous heating by the three modes balances the neutrino cooling, referred to below as the Heating = Cooling ($H=C$) curve. The $H=C$ curve can be calculated prior to carrying out an evolution using the quasi-stationary solutions for the mode amplitudes. If the spin frequency of the star upon becoming unstable is below the peak of the $H=C$ curve, then the star will reach a state of thermal equilibrium. When such a state is reached we find several possible scenarios. The star can:  (1) undergo a cyclic evolution; (2) reach a true equilibrium in which the accretion torque is balanced by the rate of loss of angular momentum via gravitational radiation; or (3) evolve in thermal equilibrium until it reaches the peak of the $H=C$ curve, which occurs on a timescale of about $10^6$ yr,  and subsequently enter a regime of thermal runaway. On the other hand, if the star cannot find a state of thermal equilibrium, then it enters a regime of thermogravitational runaway within a few hundred years of crossing the r-mode stability curve. When this happens, the r-mode amplitude increases beyond the second parametric instability, and more inertial modes would need to be included to correctly model the nonlinear effects. This will be done in a later paper.

This paper focuses on showing how nonlinear mode couplings
affect the evolution of the temperature and spin frequency of
a neutron star once it becomes prone to the r-mode CFS instability.
We do this in the context of three mode coupling, which may
be sufficient for large enough dissipation. To illustrate the
types of behavior that arise, we adopt a very specific model in
which the mode frequencies and couplings are computed for an
incompressible star, modes damp via shear viscosity, boundary layer viscosity
and hyperon bulk viscosity, and the star cools via a mixture of fast and
slow processes. This model involves several parameters that are
uncertain, and we vary these to find `phase diagrams' in which
different generic types of behavior are expected. Moreover, the
model itself is simplified: (1) A more realistic treatment
of the modes could include buoyant forces, and also mixtures
of superfluids or of superfluid and normal fluid in different
regions. (2) Dissipation rates, particularly from bulk viscosity,
depend on the composition of high density nuclear matter, which
could differ from what we assume.

Nevertheless, although the quantitative details may differ from
what we compute, we believe that many features of our calculations
ought to be robust. More sophisticated treatment of the modes
of the star will still find a dense set of modes confined to
a relatively small range of frequencies. Most importantly, this
set will exhibit numerous three mode resonances, which is the
prerequisite for strong nonlinear effects at small mode amplitudes.
Thus, whenever the unstable r-mode can pass its lowest parametric
instability threshold, it must start exciting its daughters.
Whether or not that occurs depends on the temperature dependence
of the dissipation rate of the r-mode; for the models considered
here, where bulk viscosity is relatively unimportant, soon after
the star becomes unstable its r-mode amplitude passes its
first parametric instability threshold.
Once that happens, the generic types of behavior we find -
cycles, steady states, slow and fast runaway - ought to follow
suit. The details of when different behaviors arise will depend
on the precise features of the stellar model, but the principles
we outline here (parametric instability, quasisteady evolution,
competition between heating and cooling) ought to apply quite
generally.

In Sec.\ \ref{SecEvol} we describe the evolution equations of the three modes, the angular frequency and the temperature of the neutron star.  We first show  how the equations of motion for the modes of Schenk {\it et al.\ }couple to the rotational frequency of the star in the limit of slow rotation. We then give a short review of the parametric instability threshold and the quasi-stationary solutions of the three-mode system.   The thermal and spin evolution of the star is discussed next. This is followed by a description of the driving and damping rates used. Sec.\ \ref{overview} provides an overview of the results, which includes a discussion of each evolution scenario and of the initial conditions and input physics that lead to each scenario.  Sec.\ \ref{SecCycle} discusses cyclic evolution in more detail. An evolution that leads to an equilibrium steady state is presented next in Sec.\ \ref{SecSteady}. The two types of thermal runaway are then discussed in Sec.\ \ref{SecRun}. The prospects for detecting gravitational radiation for the evolutions in which the three-mode system correctly models the nonlinear effects are considered in Sec.\ \ref{detection}. We summarize the results  in the conclusion. Appendix A sketches a derivation of the equations of motion for the three modes and Appendix B contains a stability analysis of the evolution equations around the thermal equilibrium state. 
 
\section{Evolution Equations}
\label{SecEvol}
\subsection{Three mode system: coupling to uniform rotation}
%Schenk {\it et al.} expand the Lagrangian displacement and its derivative in its basis of linear eigenvectors
%\[
%\left[
%\begin{array}{c}
%{\bf\xi}(t) \\
%{\bf \dot{\xi}}(t)
%\end{array} 
%\right]
% = \sum_A c_A(t)
%\left[ 
%\begin{array}{c}
%{\bf \xi}_A \\
%i \omega_A(t) {\bf\xi}_A 
%\end{array}
%\right] 
%\]
%which leads to equations of motion 
In this section we review the equations of motion for the three-mode system in the limit of slow rotation. In terms of rotational phase $\tau$ for the time variable with $d \tau =\Omega\;dt$ Eq.\ (2.49) of Schenk {\it et al.} \cite{Schenk} can be rewritten as 
\begin{eqnarray}
\label{Amplitudes}
\frac{d C_\alpha}{d\tau} &=& i \tilde{\omega}_\alpha C_\alpha + \frac{\gamma_\alpha}{\Omega} C_\alpha - \frac{2 i \tilde{\omega}_\alpha \tilde{\kappa}}{\sqrt{\Omega}} C_\beta C_\gamma , \\ \nonumber
\frac{d C_\beta}{d\tau} &=& i \tilde{\omega}_\beta C_\beta - \frac{\gamma_\beta}{\Omega} C_\beta - \frac{2 i \tilde{\omega}_\beta \tilde{\kappa}}{\sqrt{\Omega}} C_\alpha C_\gamma^\star , \\ \nonumber
\frac{d C_\gamma}{d\tau} &=& i \tilde{\omega}_\gamma C_\gamma - \frac{\gamma_\gamma}{\Omega} C_\gamma - \frac{2 i \tilde{\omega}_\gamma \tilde{\kappa}}{\sqrt{\Omega}} C_\alpha C_\beta^\star .
\end{eqnarray}
Here the scaled frequency $\tilde{\omega}_j$ is defined to be $\tilde{\omega}_j = \omega_j/\Omega$, the dissipation rates of the daughter modes are $\gamma_\beta$ and $\gamma_\gamma$, $\gamma_\alpha$ is the sum of the driving and damping rates of the r-mode  $\gamma_\alpha = \gamma_{GR} - \gamma_{\alpha\, v}$, and the dimensionless coupling is $\tilde{\kappa} = \kappa/(M R^2 \Omega^2)$.  These amplitude variables are complex and can be written in terms of the variables of Ref.\  \cite{Schenk} as $C_j(t) = \sqrt{\Omega(t)} c_j(t)$ (see Appendix A for a derivation of Eqs.\ \ref{Amplitudes}). The index $j$ loops over the three modes $j = \alpha, \beta, \gamma$, where $\alpha$ labels the r-mode or parent mode and $\beta$ and $\gamma$ label the two daughter modes in the mode triplet. 

When the daughter mode amplitudes are much smaller than that of the parent mode, one can approximate the parent mode amplitude as constant. Under this assumption one performs a linear stability analysis on Eqs. (\ref{Amplitudes}) and finds the r-mode amplitude when the two daughter modes become unstable (see Eqs. (B5-B7) of Ref.  \cite{Jeandrew3} for a full derivation). This amplitude is the parametric instability threshold
\begin{eqnarray}
|C_\alpha|^2 = \frac{\gamma_\beta \gamma_\gamma}{4 \tilde{\kappa}^2 \tilde{\omega}_\beta \tilde{\omega}_\gamma \Omega} \left[1 + \Omega^2 \left(\frac{\delta \tilde{\omega}}{\gamma_\beta + \gamma_\gamma} \right)^2 \right] ,
\label{threshold}
\end{eqnarray} 
where the fractional detuning is $\delta \tilde{\omega} =  \tilde{\omega}_\alpha - \tilde{\omega}_\beta -\tilde{\omega}_\gamma$.
Thorough explorations of the phase space of damped three-mode systems were performed by Dimant \cite{Dimant} and Wersinger {\it et al.} \cite{Wer}. 
 
 For the three modes at the lowest parametric instability threshold, $\tilde{\omega}_\alpha \approx 0.66$,  $\tilde{\omega}_\beta \approx 0.44$, $\tilde{\omega}_\gamma \approx 0.22$, $\tilde{\kappa} \approx 0.19$ and $|\delta\tilde{\omega}| \approx 3.82 \times 10^{-6}$. Note that $\tilde\omega$ is twice the $w$ of Brink {\it et al.} \cite{Jeandrew1, Jeandrew2, Jeandrew3}. Here $\beta$ labels the mode with $n=13, m=-3$ and $\gamma$ labels the $n=14, m=1$ mode. The amplitude the r-mode has to reach before exciting these two daughter modes is $|C_\alpha| \approx 1.5 \times 10^{-5} \sqrt{\Omega}$ \cite{Jeandrew3}. 
 
We next rescale the rotational phase $\tau$ by the fractional detuning as $\tilde{\tau} = \tau |\delta \tilde{\omega}|$ and the mode amplitudes by 
\begin{eqnarray}
\label{norm}
|C_\alpha|_0 &=& \frac{|\delta \tilde{\omega}| \sqrt{\Omega_c}}{4 \tilde{\kappa} \sqrt{\tilde{\omega}_\beta \tilde{\omega}_\gamma}}, \; |C_\beta|_0 = \frac{|\delta \tilde{\omega}| \sqrt{\Omega_c}}{4 \tilde{\kappa} \sqrt{\tilde{\omega}_\alpha \tilde{\omega}_\gamma}}, \\ \nonumber
 |C_\gamma|_0 &=& \frac{|\delta \tilde{\omega}| \sqrt{\Omega_c}}{4 \tilde{\kappa} \sqrt{\tilde{\omega}_\beta \tilde{\omega}_\alpha}} \;,
\end{eqnarray}
which for the r-mode is, up to a factor of $\sqrt{\tilde{\Omega}} = \sqrt{\Omega/\Omega_c}$, the no-damping limit of the parametric instability threshold below which no oscillations will occur. 
The coupled equations become
\begin{eqnarray}
\label{eqcode}
\frac{d \bar{C}_\alpha}{d\tilde{\tau}} &=& \frac{i \tilde{\omega}_\alpha}{|\delta \tilde{\omega}|} \bar{C}_\alpha + \frac{\tilde{\gamma}_\alpha}{|\delta \tilde{\omega}| \tilde{\Omega}} \bar{C}_\alpha - \frac{i}{2 \sqrt{\tilde{\Omega}}} \bar{C}_\beta \bar{C}_\gamma ,  \\ \nonumber
\frac{d \bar{C}_\beta}{d \tilde{\tau}} &=& \frac{i \tilde{\omega}_\beta}{|\delta \tilde{\omega}|} \bar{C}_\beta - \frac{\tilde{\gamma}_\beta}{|\delta \tilde{\omega}| \tilde{\Omega}} \bar{C}_\beta - \frac{i}{2 \sqrt{\tilde{\Omega}}} \bar{C}_\alpha \bar{C}_\gamma^\star , \\ \nonumber
\frac{d \bar{C}_\gamma}{d\tilde{\tau}}&=&\frac{i \tilde{\omega}_\gamma}{|\delta \tilde{\omega}|} \bar{C}_\gamma - \frac{\tilde{\gamma}_\gamma}{|\delta \tilde{\omega}| \tilde{\Omega}} \bar{C}_\gamma - \frac{i}{2 \sqrt{\tilde{\Omega}}} \bar{C}_\alpha \bar{C}_\beta^\star  ,
\end{eqnarray}
with $\bar{C}_j  = C_j/|C_j|_0$ and $\tilde{\gamma}_j = \gamma_j/\Omega_c$ being the newly rescaled amplitudes and dissipation/driving rates, respectively.
  
\subsubsection{Quasi-Stationary Solution}

In terms of amplitudes and phase variables $C_j = |C_j| e^{i \phi_j}$ Eqs.\ (\ref{eqcode}) can be rewritten as
\begin{eqnarray}
\label{phase}
\frac{d |\bar{C}_\alpha|}{d\tilde{\tau}} &=&\frac{ \tilde{\gamma}_\alpha}{\tilde{\Omega} |\delta \tilde{w}|} |\bar{C}_\alpha| - \frac{\sin \phi |\bar{C}_\beta| |\bar{C}_\gamma|}{2 \sqrt{\tilde{\Omega}}} ,  \\  \nonumber
\frac{d |\bar{C}_\beta|}{d\tilde{\tau}} &=& - \frac{ \tilde{\gamma}_\beta}{\tilde{\Omega} |\delta \tilde{w}|} |\bar{C}_\beta| + \frac{\sin \phi |\bar{C}_\alpha| |\bar{C}_\gamma|}{2 \sqrt{\tilde{\Omega}}} , \\  \nonumber
\frac{d |\bar{C}_\gamma|}{d\tilde{\tau}} &=& - \frac{ \tilde{\gamma}_\gamma}{\tilde{\Omega} |\delta \tilde{w}|} |\bar{C}_\gamma| +\frac{\sin \phi |\bar{C}_\alpha| |\bar{C}_\beta|}{2 \sqrt{\tilde{\Omega}}} , \\  \nonumber
\frac{d \phi}{d \tilde{\tau}} &=&  \frac{\delta \tilde{\omega}}{|\delta \tilde{\omega}|}  - \frac{\cos\phi}{2 \sqrt{\tilde \Omega}} \left(\frac{|\bar{C}_\beta| |\bar{C}_\gamma|}{|\bar{C}_\alpha|}-\frac{|\bar{C}_\alpha| |\bar{C}_\gamma|}{|\bar{C}_\beta|} - \frac{|\bar{C}_\beta| |\bar{C}_\alpha|}{|\bar{C}_\gamma|} \right) ,
\end{eqnarray}
where we have defined the relative phase difference as $\phi = \phi_\alpha - \phi_\beta - \phi_\gamma$. 
These equations have the stationary solution
\begin{eqnarray}
\label{stationarySol}
|\bar{C}_\alpha|^2 = \frac{4 \tilde{\gamma}_\beta \tilde{\gamma}_\gamma}{\tilde{\Omega} |\delta \tilde{\omega}|^2} \left(1 + \frac{1}{\tan^2 \phi}\right) , \\ \nonumber
|\bar{C}_\beta|^2 = \frac{4 \tilde{\gamma}_\alpha \tilde{\gamma}_\gamma}{\tilde{\Omega} |\delta \tilde{\omega}|^2} \left(1 + \frac{1}{\tan^2 \phi}\right) , \\ \nonumber
|\bar{C}_\gamma|^2 = \frac{4 \tilde{\gamma}_\alpha \tilde{\gamma}_\beta}{\tilde{\Omega} |\delta \tilde{\omega}|^2} \left(1 + \frac{1}{\tan^2 \phi}\right) , \\ \nonumber
\tan \phi = \frac{\tilde{\gamma}_\beta + \tilde{\gamma}_\gamma - \tilde{\gamma}_\alpha}{\tilde{\Omega} |\delta \tilde{\omega}|} .
\end{eqnarray}
Note that in the limit in which $\gamma_\beta + \gamma_\gamma >> \gamma_\alpha$ the stationary solution for the r-mode amplitude $|C_\alpha|$ is the same as the parametric instability threshold.
%\begin{eqnarray}
%|C_\alpha|^2 = \frac{\gamma_\beta \gamma_\gamma \tilde{\Omega}}{4 \tilde{\kappa}^2 \tilde{\omega}_\beta \tilde{\omega}_\gamma} \left(\frac{\delta \omega}{\gamma_\beta + \gamma_\gamma} \right)^2 
%\end{eqnarray}

\subsection{Temperature and Spin Evolution}
%The heat equation for regular time units [time in seconds] is
%\begin{equation}
%C(T) \frac{dT}{dt} = 2 E_c F_v(T) +  K_n \dot{M} c^2 - L_\nu(T)
%\end{equation}
%For $d\tilde{\tau} = 2 \Omega |\delta \tilde{\omega}| dt$ 
%\begin{equation}
%C(T) \frac{dT}{d\tilde{\tau}} = 2 E_c \frac{\tilde{F}_v(T)}{\tilde{\Omega} |\delta \tilde{\omega}|} +  \frac{K_n \dot{M} c^2 - L_\nu(T)}{2 \Omega_c \tilde{\Omega} |\delta \tilde{\omega}|}
%\end{equation}
%\begin{equation}
%C(T) \frac{dT}{d\tilde{\tau}} = 2 M R^2 \Omega_c^2 \frac{(\tilde{\gamma}_{\alpha,v} |c_\alpha|^2 + \tilde{\gamma}_\beta |c_\beta|^2 + \tilde{\gamma}_\gamma |c_\gamma|^2))}{|\delta \tilde{\omega}|} +  \frac{K_n \dot{M} c^2 - L_\nu(T)}{2 \Omega_c \tilde{\Omega} |\delta \tilde{\omega}|}
%\end{equation}
The spin evolution equation is obtained from conservation of total angular momentum $J$, where
 \begin{equation}
J = I \Omega + J_{\rm{phys}} .
\label{ang0}
\end{equation}
Following Eq (K39-K42) of Schenk {\it et al.} \cite{Schenk} the physical angular momentum of the perturbation can be written as
\begin{eqnarray}
\label{ang}
\Omega J_{\rm{phys}} &= &\sum_{A B} C_B^{\star} C_A \int d^3x \rho [(\hat{\Omega} \times \xi_B^{\star})\cdot (\hat{\Omega} \times \xi_A)  \\ \nonumber
&-& i \frac{(\tilde{\omega}_A+\tilde{\omega}_B)}{2} \xi_B^{\star} \cdot (\hat{\Omega} \times \xi_A)].
\end{eqnarray}
Since the eigenvectors $\xi_A \propto e^{i m_A \phi}$ the cross-terms will vanish for modes with different magnetic quantum numbers $m$ as $\int e^{i (m_A-m_B) \phi} d\phi = 0$ for $m_A \ne m_B$. Eq. (\ref{ang}) can be re-written for our triplet of modes as
\begin{eqnarray}
J_{\rm{phys}} = M R^2 (k_{\alpha \alpha} |C_\alpha|^2  + k_{\beta \beta} |C_\beta|^2+ k_{\gamma \gamma} |C_\gamma|^2) ,
\label{ang3}
\end{eqnarray}
where $k_{\alpha \alpha}$ is defined as
\begin{equation}
k_{\alpha \alpha} =\frac{1}{M R^2} \int d^3x \rho [(\hat{\Omega} \times \xi_\alpha^{\star})\cdot (\hat{\Omega} \times \xi_\alpha) - i \tilde{\omega}_\alpha \xi_\alpha^{\star} \cdot (\hat{\Omega} \times \xi_\alpha)]
\end{equation}
and similarly for $k_{\beta \beta}$ and $k_{\gamma \gamma}$.
In terms of the scaled variables $\bar{C}_j  = C_j/|C_j|_0$ (with $|C_j|_0$ defined in Eq.\ (\ref{norm})) the angular momentum of the perturbation can be written as
\begin{eqnarray}
J_{\rm{phys}} = \frac{M R^2 \Omega_c |\delta \tilde{\omega}|^2}{(4 \tilde{k})^2 \tilde{\omega}_\alpha \tilde{\omega}_\beta \tilde{\omega}_\gamma} (k_{\alpha \alpha} |\bar{C}_\alpha|^2 \tilde{\omega}_\alpha \\ \nonumber
+ k_{\beta \beta} |\bar{C}_\beta|^2 \tilde{\omega}_\beta+ k_{\gamma \gamma} \tilde{\omega}_\gamma |\bar{C}_\gamma|^2) .
\end{eqnarray}
We chose the same normalization for the eigenfuctions as Refs.\ \cite{Schenk,arras,Jeandrew1, Jeandrew2,Jeandrew3} so that at unit amplitude all modes have the same energy $\epsilon_\alpha = M R^2 \Omega^2$. The energy of a mode $\alpha$ is $E_\alpha = M R^2 \Omega^2 |c_\alpha|^2 = M R^2 \Omega |C_\alpha|^2$. The rotating frame energy is the same as the canonical energy and physical energy \cite{Schenk}.  The canonical angular momentum and the canonical energy of the perturbation satisfy the general relation $E_c = - (\omega/m) J_c$ \cite{FS}. 

Angular momentum is gained because of accretion and lost via gravitational waves emission
\begin{equation}
\label{consAng}
\frac{dJ}{dt} = 2  \gamma_{GR} J_{c\;\rm{rmode}} +  \dot{M} \sqrt{GMR},
\end{equation}
where $J_{c\;\rm{rmode}} = - (m_\alpha/\omega_{\alpha}) \epsilon_\alpha |c_\alpha|^2 = - 3 M R^2 \Omega |c_\alpha|^2 = - 3 M R^2 |C_\alpha|^2$.  Eq.\ (\ref{consAng}) can be rewritten in terms of the scaled variables $\bar{C}_j$ as
\begin{equation}
\label{ang2}
\frac{dJ}{d\tilde{\tau}} = -  \frac{6 \tilde{\gamma}_{\rm{GR}}}{\tilde{\Omega}} \frac{M R^2 \Omega_c |\delta \tilde{\omega}|}{(4 \tilde{k})^2 \tilde{\omega}_\beta \tilde{\omega}_\gamma}  |\bar{C}_\alpha|^2  + \frac{\dot{M} \sqrt{G M R}}{\Omega_c \tilde{\Omega} |\delta \tilde{\omega}|} .
\end{equation}
Thermal energy conservation gives the temperature evolution equation
\begin{eqnarray}
\label{consE}
C(T) \frac{dT}{dt} &=& \sum_j 2 E_j  \gamma_j  + K_n \dot{M} c^2 - L_\nu(T), \\ \nonumber
&=& 2 M R^2 \Omega (\gamma_{\alpha\,v} |C_\alpha|^2 + \gamma_\beta |C_\beta|^2 \\ \nonumber
 &+& \gamma_\gamma |C_\gamma|^2)  + K_n \dot{M} c^2 - L_\nu(T).
\end{eqnarray} 
The three terms on the right hand side of the equation represent viscous heating, nuclear heating and
neutrino cooling. The specific heat is taken to be $C(T) \approx 1.5 \times 10^{38} \; T_8 \; \rm{erg \; K}^{-1}$, where $T=T_8 \times 10^8$ K. Nuclear heating occurs because of pycnonuclear reactions and neutron emission in the inner crust \cite{Brown}. At large accretion rates such as that of the brightest LMXBs of $\dot{M} \approx 10^{-8} M_\odot/\rm{yr}$, the accreted helium and hydrogen burns stably and most of the heat released in the crust is conducted into the core of the neutron star, where neutrino emission is assumed to regulate the temperature of the star \cite{Brown, Schatz}. The nuclear heating constant is taken to be $K_n \approx 1 \times 10^{-3}$ \cite{Brown}. Following Ref. \cite{Wagoner}, we take the neutrino luminosity to be
\begin{eqnarray}
L_\nu &=& L_{\rm{dU}} T_8^6 R_{\rm{dU}}(T/T_p) + L_{\rm{mU}} T_8^8 R_{\rm{mU}}(T/T_p) \\ \nonumber
 &+& L_{\rm{e-i}} T_8^6 + L_{\rm{n-n}} T_8^8 + L_{\rm{Cp}} T_8^7 ,
\end{eqnarray}
where  the constants for the modified and direct URCA reactions are defined by $L_{\rm{mU}} = 1.0 \times 10^{32} \; \rm{erg\; sec}^{-1}$, $L_{\rm{dU}} = f_{\rm{dU}} 
\times 10^8 L_{\rm{mU}}$ \cite{Yakovlev, Yakovlev2}, and the electron-ion, neutron-neutron neutrino bremsstrahlung and Cooper pairing of neutrons are given by  $L_{\rm{e-i}} = 9.1 \times 10^{29}\;\rm{erg\; sec}^{-1}$ \cite{Brown},  $L_{\rm{n-n}} \approx 0.01 L_{\rm{mU}}$, $L_{\rm{Cp}} = 8.9 \times 10^{31}\; \rm{erg\; sec}^{-1}$ \cite{Yakovlev3}. The fraction of the star $f_{\rm{dU}}$ that is above the density threshold for direct URCA reactions is in general dependent on the equation of state \cite{YP} and in this work we treat $f_{\rm{dU}}$ a free parameter with values between 0 and 1.

The proton superfluid reduction factors for the modified and direct URCA reactions are taken from Ref. \cite{Yakovlev2} (see Eqs. (32) and (51) in Ref. \cite{Yakovlev2}):
\begin{eqnarray}
R_{\rm{dU}}(T/T_p) &=&\left[0.2312 + \sqrt{(0.76880)^2 + (0.1438 v)^2}\right]^{5.5} \;\;\; \\ \nonumber
 &\times& \exp\left(3.427- \sqrt{(3.427)^2 + v^2}\right) , \\ \nonumber
R_{\rm{mU}}(T/T_p) &=& \left(0.2414 + \sqrt{(0.7586)^2+(0.1318 v)^2}\right)^7 , \\ \nonumber
&\times&\exp\left(5.339-\sqrt{(5.339)^2+(2 v)^2}\right)
\end{eqnarray}
where  the dimensionless gap amplitude $v$ for the singlet type superfluidity is given by
%\begin{eqnarray}
%a=0.1477+\sqrt{(0.8523)^2+(0.1175 v)^2} \\ \nonumber
%b=0.1477+\sqrt{(0.8523)^2+(0.1297 v)^2}
%\end{eqnarray}
\begin{equation}
v=\sqrt{1-\frac{T}{T_p}} \left(1.456-0.157 \sqrt{\frac{T_p}{T}} + 1.764 \frac{T_p}{T} \right).
\end{equation}
Similar to Ref.\ \cite{Wagoner}, we use $T_p = 5.0 \times 10^9$ K.
In terms of the scaled variables Eq.\ (\ref{consE}) becomes
\begin{eqnarray}
\label{thermalevol}
C(T) \frac{dT}{d\tilde{\tau}} = \frac{2 M R^2 \Omega_c^2 |\delta \tilde{\omega}|}{(4 \tilde{\kappa})^2 \tilde{\omega}_\alpha  \tilde{\omega}_\beta  \tilde{\omega}_\gamma} (\tilde{\omega}_\alpha \tilde{\gamma}_{\alpha\,v} |\bar{C}_\alpha|^2 + \tilde{\omega}_\beta \tilde{\gamma}_\beta |\bar{C}_\beta|^2  \\ \nonumber
+ \tilde{\omega}_\gamma \tilde{\gamma}_\gamma |\bar{C}_\gamma|^2)
+  \frac{K_n \dot{M} c^2 - L_\nu(T)}{ \Omega_c \tilde{\Omega} |\delta \tilde{\omega}|}.
\end{eqnarray}
\subsection{Temperature and Spin Evolution with the Mode Amplitudes in Quasi-Stationary States}
Assuming that the amplitudes evolve through a series of spin- and temperature-dependent steady states, i.e., $dC_i/d\tilde{\tau} \approx 0$, the spin and thermal evolution equations can be rewritten by taking $J\approx I \Omega$ and using Eqs.\ (\ref{stationarySol}) in Eq.\ (\ref{ang2}).
\begin{eqnarray}
\label{spinSteady}
\frac{d \tilde{\Omega}}{d\tilde{\tau}} = - \frac{6 \tilde{\gamma}_{\rm{GR}}}{\tilde{\Omega}^2 |\delta \tilde{\omega}|} \frac{\tilde{\gamma}_\beta \tilde{\gamma}_\gamma}{4 \tilde{k}^2\tilde{I}  \tilde{\omega}_\beta \tilde{\omega}_\gamma} k_{\alpha \alpha} \left(1 + \frac{1}{\tan^2 \phi}\right) \\ \nonumber
  +  \frac{\dot{M}}{\Omega_c^2} \frac{\sqrt{G M R}}{M R^2 \tilde{I} \tilde{\Omega} |\delta \tilde{\omega}|} ,
\end{eqnarray}
where $\tilde{I} = I/(M R^2)$.
The thermal evolution of the system is given by
\begin{eqnarray}
\label{thermalSteady}
C(T) \frac{dT}{d\tilde{\tau}} =\frac{2 M R^2 \Omega_c^2}{(4 \tilde{\kappa})^2 \tilde{\omega}_\alpha  \tilde{\omega}_\beta  \tilde{\omega}_\gamma} \frac{\tilde{\gamma}_\alpha \tilde{\gamma}_\beta \tilde{\gamma}_\gamma}{\tilde{\Omega}  |\delta \tilde{\omega}|} \left(\frac{\tilde{\omega}_\alpha \tilde{\gamma}_{\alpha,v}}{\tilde{\gamma}_\alpha}+ \tilde{\omega}_\beta  \right. \\ \nonumber \left. +
\tilde{\omega}_\gamma \right) \left(1+\frac{1}{\tan^2\phi}\right)+  \frac{K_n \dot{M} c^2 - L_\nu(T)}{\Omega_c \tilde{\Omega} |\delta \tilde{\omega}|}.
\end{eqnarray}
By setting the right hand side of the above equation to zero, one can find the Heating = Cooling ($H=C$) curve. Below, we find that Eqs.\ (\ref{spinSteady})-(\ref{thermalSteady}) describe the evolution very well throughout the unstable regime. These equations are a minimal physical model for the effects of nonlinear coupling on r-mode evolution. 
\subsection{Sources of Driving and Dissipation}
\label{dissipation}
The damping mechanisms are shear viscosity, boundary layer viscosity and hyperon bulk viscosity; for modes $j=\alpha, \beta, \gamma$ we write
\begin{equation}
\gamma_{j \; v}(\Omega,T) =  \gamma_{j \; \rm{sh}}(T) +
\gamma_{j\; \rm{bl}}(\Omega,T)  + \gamma_{j\; \rm{hb}}(\Omega,T).
\end{equation}  
The r-mode is driven by gravitational radiation and damped by these dissipation mechanisms, while the 
pair of daughter modes  ($n=13,m=-3$ labeled as $\beta$ and $n=14, m=1$ labeled as $\gamma$)  is affected only by the viscous damping.  Brink {\it et al.} \cite{Jeandrew1, Jeandrew2, Jeandrew3} determined that this pair of modes is excited at the lowest parametric instability threshold. Their model uses the Bryan \cite{Bryan} modes of an incompressible star, which has the advantage that the mode eigenfrequencies (and eigenfunctions) are known analytically. This enables them to find near resonances efficiently. We are using their results, but we include more realistic effects such as bulk viscosity, whose effect vanishes in the incompressible limit ($\Gamma_1 \to \infty$ in Eq.\ (\ref{bulkEnergy}))

For our benchmark calculations, we adopt the neutron star model of Owen {\it et al.} Ref. \cite{LOM}  ($n=1$ polytrope, $M = 1.4 M_\odot$, $\Omega_c = 8.4 \times 10^3 \; \rm{rad} \; \rm{sec}^{-1}$ and $R = 12.53$ km) and use their gravitational driving rate and shear viscous damping rate for the r-mode 

\begin{eqnarray}
\gamma_{GR}(\Omega) &\simeq& \frac{\tilde{\Omega}^6}{3.26} \; \rm{sec}^{-1} , \\ \nonumber
\gamma_{\alpha \; \rm{sh}}(T) &\simeq& \frac{1}{\tau_{\rm{sh}}} \frac{1}{T_8^2} ,
\end{eqnarray}
where $\tau_{\rm{sh}} = 2.56 \times 10^6\; \rm{sec}$. (In Sec.\ \ref{detection} we consider approximate scalings with M and R.)

The damping rate due to shear viscosity for the two daughter modes is calculated using the Bryan modes for a star with the same mass and radius 
\begin{eqnarray}
\gamma_{\beta\; \rm{sh}}(T) &\simeq& 3.48 \times 10^{-4} \; \rm{sec}^{-1}  \frac{1}{T_8^2} , \\ \nonumber
\gamma_{\gamma\; \rm{sh}}(T) &\simeq& 4.52 \times 10^{-4} \; \rm{sec}^{-1} \frac{1}{T_8^2} .
\end{eqnarray}
 %We have used the shear viscosity coefficient $\eta$ for neutron-neutron scattering proposed by Ref. \cite{Curt}. 
 The geometric contribution $\gamma_{\rm{sh}}/\eta$ of the individual modes increases significantly with the degree $n$ of the mode scaling approximatively like $n^3$  for large $n$ (see Eq. (29) of Brink {\it et al.} \cite{Jeandrew2} for an analytic fit to the shear damping rates computed for the 5,000 modes in their network), and hence the inertial modes with $n=13$ and $n=14$ have shear damping rates about three orders of magnitude larger than that of the r-mode.

The damping due to boundary layer viscosity is calculated using Eq.\ (4) of Ref. \cite{BU},
\begin{eqnarray}
\gamma_{\alpha \; \rm{bl}}(T,\Omega) &\simeq& 0.009\; {\rm sec^{-1}} \; S_{\rm{ns}}^2 \frac{\sqrt{\tilde{\Omega}}}{T_8} , \\ \nonumber
\gamma_{\beta \; \rm{bl}}(T,\Omega) &\simeq& 0.028\; {\rm sec^{-1}} \;  S_{\rm{ns}}^2 \frac{\sqrt{\tilde{\Omega}}}{T_8} , \\ \nonumber
\gamma_{\gamma \; \rm{bl}}(T,\Omega) &\simeq& 0.021\; {\rm sec^{-1}} S_{\rm{ns}}^2 \frac{\sqrt{\tilde{\Omega}}}{T_8} .
\end{eqnarray}
%\begin{eqnarray}
%\gamma_{\alpha \; \rm{bl}}(T,\Omega) &\simeq& 0.0013\; {\rm sec^{-1}} \; \frac{R_6^2 \tilde{\omega}_\alpha^{5/2}}{\tilde{M}} S_{\rm{ns}}^2 \frac{\sqrt{\Omega}}{T_8} \\ \nonumber
%\gamma_{\beta \; \rm{bl}}(T,\Omega) &\simeq& 0.012\; {\rm sec^{-1}} \; \frac{R_6^2 \tilde{\omega}_\beta^{5/2}}{\tilde{M}} S_{\rm{ns}}^2 \frac{\sqrt{\Omega}}{T_8} \\ \nonumber
%\gamma_{\gamma \; \rm{bl}}(T,\Omega) &\simeq& 0.048\; {\rm sec^{-1}} \; \frac{R_6^2 \tilde{\omega}_\gamma^{5/2}}{\tilde{M}} S_{\rm{ns}}^2 \frac{\sqrt{\Omega}}{T_8}
%\end{eqnarray}
Analogous to Wagoner \cite{Wagoner}, we  allow the slippage factor $S_{\rm{ns}}$ to vary. The slippage factor is defined by Refs. \cite{Wagoner, LU, KM} to be $S_{\rm{ns}}^2 = (2 S_{\rm{n}}^2 +S_{\rm{s}}^2)/3$, with $S_n$ being the fractional difference in velocity of the normal fluid between the crust and the core \cite{LU} and $S_{\rm{s}}$ the fractional degree of pinning of the vortices in the crust \cite{KM}. Note that $\gamma_{\beta \; \rm{bl}}$ and $\gamma_{\gamma \; \rm{bl}}$ are both greater than $ 2 \times \gamma_{\alpha \; \rm{bl}}$ and can easily be comparable to $\gamma_{\rm{GR}}$ in the unstable regime. 

%A deficiency of the incompressible model is that $\Delta p =0$  for the r-mode (and all $n=m-1$ inertial modes), which makes the bulk viscosity coefficient zero. We account for this deficiency by using the  damping coefficient computed by  an $n=1$ polytrope.
The damping rate due to bulk viscosity produced by out-of-equilibrium hyperon reactions for the r-mode is found by fitting the results of Nayyar and Owen \cite{mohit}. This rate is taken to have a form similar to that taken by Wagoner \cite{Wagoner}
\begin{equation}
\gamma_{\alpha \; \rm{hb}} = f_{\rm{hb}} \frac{t_{0\alpha}^{-2}  \tau(T) \tilde{\Omega}^4}{1+(\tilde{\omega}_\alpha \Omega \tau(T))^2} ,
\end{equation}
and for the daughter modes
\begin{equation}
\gamma_{\beta \; \rm{hb}} = f_{\rm{hb}} \frac{t_{0\beta}^{-2}  \tau(T) \tilde{\omega}_\beta^2}{1+(\tilde{\omega}_\beta \Omega \tau(T))^2} ,
\end{equation}
and similarly foe $\gamma_{\gamma \; \rm{hb}}$. The relaxation timescale
\begin{equation}
\tau(T) = \frac{t_1 T_8^{-2}}{R_{\rm{hb}}(T/T_c)}
\end{equation}
The reduction factor is taken to be the product of two single-particle reduction factors \cite{HLY,mohit}
\begin{equation}
R_{\rm{hb}\;\rm{single}}(T/T_c) = \frac{a^{5/4} + b^{1/2}}{2} \exp\left(0.5068 - \sqrt{0.5068^2+y^2}\right)
\end{equation}
where $a = 1+ 0.3118 y^2$,  $b = 1 + 2.556 y^2$ and $y = \sqrt{1.0 - T/T_c} (1.456 - 0.157 \sqrt{T_c/T} + 1.764 T_c/T)$.
 The constants $t_1 \approx 10^{-4}  \; \rm{sec}$ and $t_{0\alpha} \approx 0.00058 \; \rm{sec}$ are found by fitting the results of Ref.\ \cite{mohit}. The factor $f_{\rm{hb}}$ allows for physical uncertainties; we take $f_{\rm{hb}}=1$ throughout the body of the paper since $T_c$ , which enters $\gamma_{j\;\rm{hb}}$ exponentially, is also uncertain.   For the daughter modes, the dissipation energy due to bulk viscosity is calculated using the modes for the incompressible star. In the slow rotation limit, it is given to leading order in $\Gamma_1^{-2}$ by
 %(this approximation has been proposed by Cutter and Lindblom \cite{CL} and adopted by Kokkotas and Stergioulas \cite{KS} for the r-mode.)}
 \begin{equation}
 \label{bulkEnergy}
 - \dot{E}_{B\,j} = \left(\frac{\zeta \omega_j^2}{\Gamma_1^2} \right) \int d^3x \left|\frac{{\bf\xi}_j \cdot \nabla p}{p} \right|^2 .
 \end{equation}
 This approximation was proposed by Cutler and Lindblom \cite{CL} and adopted by Kokkotas and Stergioulas \cite{KS} for the r-mode and by Brink {\it et al.} \cite{Jeandrew2} for the inertial modes.
 The adiabatic index $\Gamma_1$ is regarded as a parameter; we use $\Gamma_1 \approx 2$. 
The damping rate is 
\begin{equation}
\gamma_{j \, \rm{hb}} = - \frac{\dot{E}_{B\,j}}{\epsilon} ,
\end{equation}
where $\epsilon = M R^2 \Omega^2$ is the mode's energy in the rotating frame at unit amplitude and $j = \beta, \gamma$.
Using this procedure, we calculate
\begin{eqnarray}
t_{0\beta} \approx 1.4 \times 10^{-5} \; \rm{sec} , \\ \nonumber
t_{0\gamma} \approx 1.0 \times 10^{-5} \; \rm{sec} .
\end{eqnarray} 
\section{Summary of Results}
\label{overview}
 \begin{figure}
\begin{center}
\leavevmode
\epsfxsize=250pt
\epsfbox{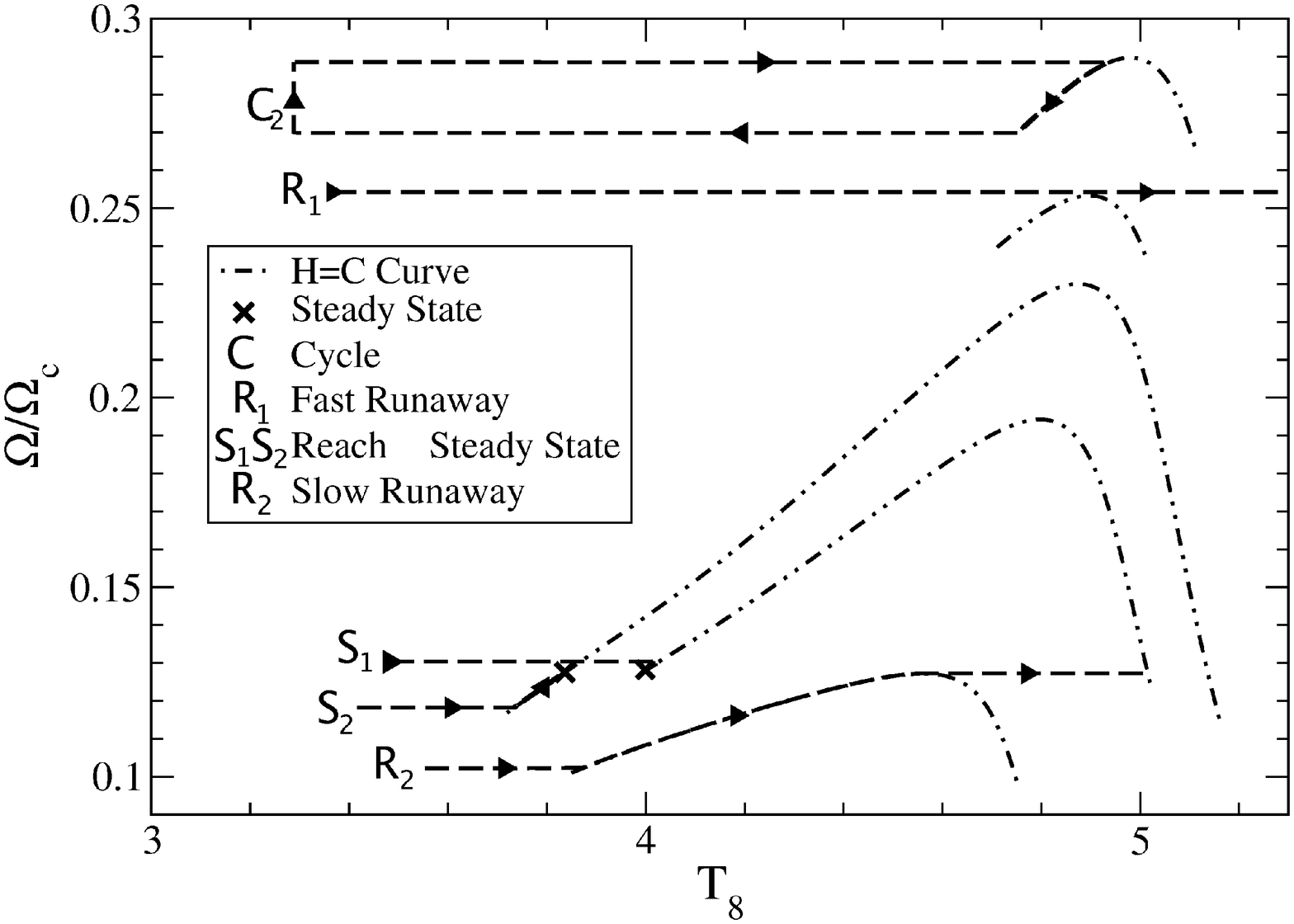}
\epsfxsize=250pt
\epsfbox{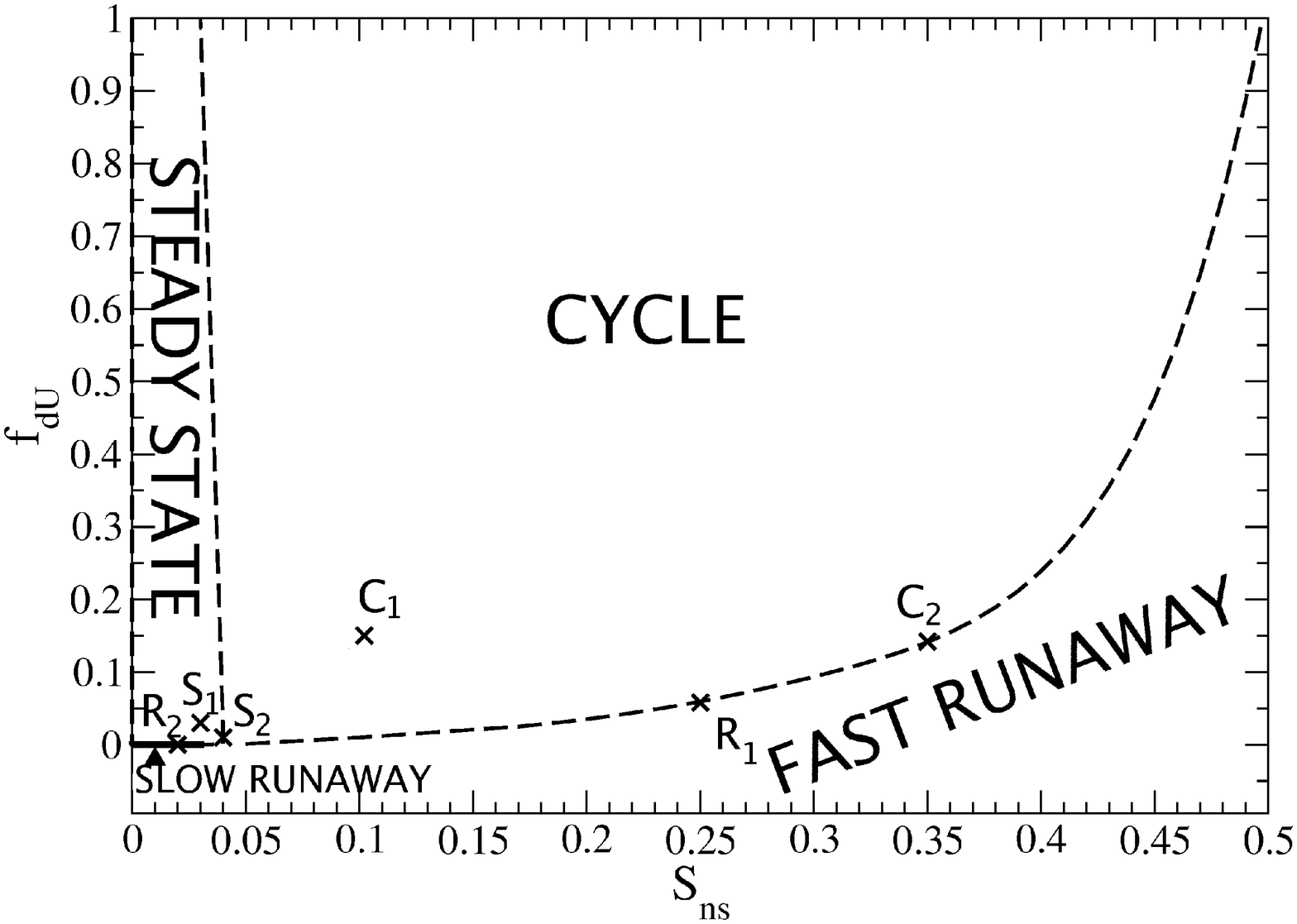}
\caption{(a)Typical trajectories for the four  observed evolution scenarios  are shown in the  $\tilde{\Omega}$ - $T_8$ phase space, where $\tilde{\Omega} = \Omega/\Omega_c$. The dashed lines ($H=C$ curves) represent the points in the $\tilde{\Omega}-T_8$ phase space where the dissipative effects of the heating from the three-modes exactly compensate the neutrino cooling for the given set of parameters ($S_{\rm{ns}}$, $f_{\rm{dU}}$, $T_c$, ...) of each evolution. 
(b)The corresponding stability regions for which these scenarios occur are plotted at fixed hyperon superfluidity temperature $T_c = 5.0 \times 10^9$ K, while varying  $f_{\rm{dU}}$ and $S_{\rm{ns}}$. The position of the initial angular velocity and temperature ($\tilde{\Omega}_{\rm{in}}$, $T_{8\;\rm{in}}$) with respect to the maximum of this curve determines the stability of the evolution. (I)  $\tilde{\Omega}_{\rm{in}} > \tilde{\Omega}_{H=C \; \rm{max}}$. Trajectory $R_1$.  Fast Runaway Region. After the r-mode becomes unstable the star heats up, does not find a thermal equilibrium state and continues heating up until a thermogravitational runaway occurs. (II) $\tilde{\Omega}_{\rm{in}} < \tilde{\Omega}_{H=C \; \rm{max}}$. The evolutions are either stable or, if there is a runaway, it  occurs on timescales comparable to the accretion timescale. The possible trajectories are (1)Trajectory C.  Cycle Region.  (2) Trajectories $S_1$ and $S_2$. Steady State Region. (3) Trajectory $R_2$. Slow Runaway Region. }
 \renewcommand{\arraystretch}{0.75}
 \renewcommand{\topfraction}{0.6}
\label{scenario}
\end{center}
\end{figure}
\begin{figure}
\begin{center}
\leavevmode
\epsfxsize=240pt
\epsfbox{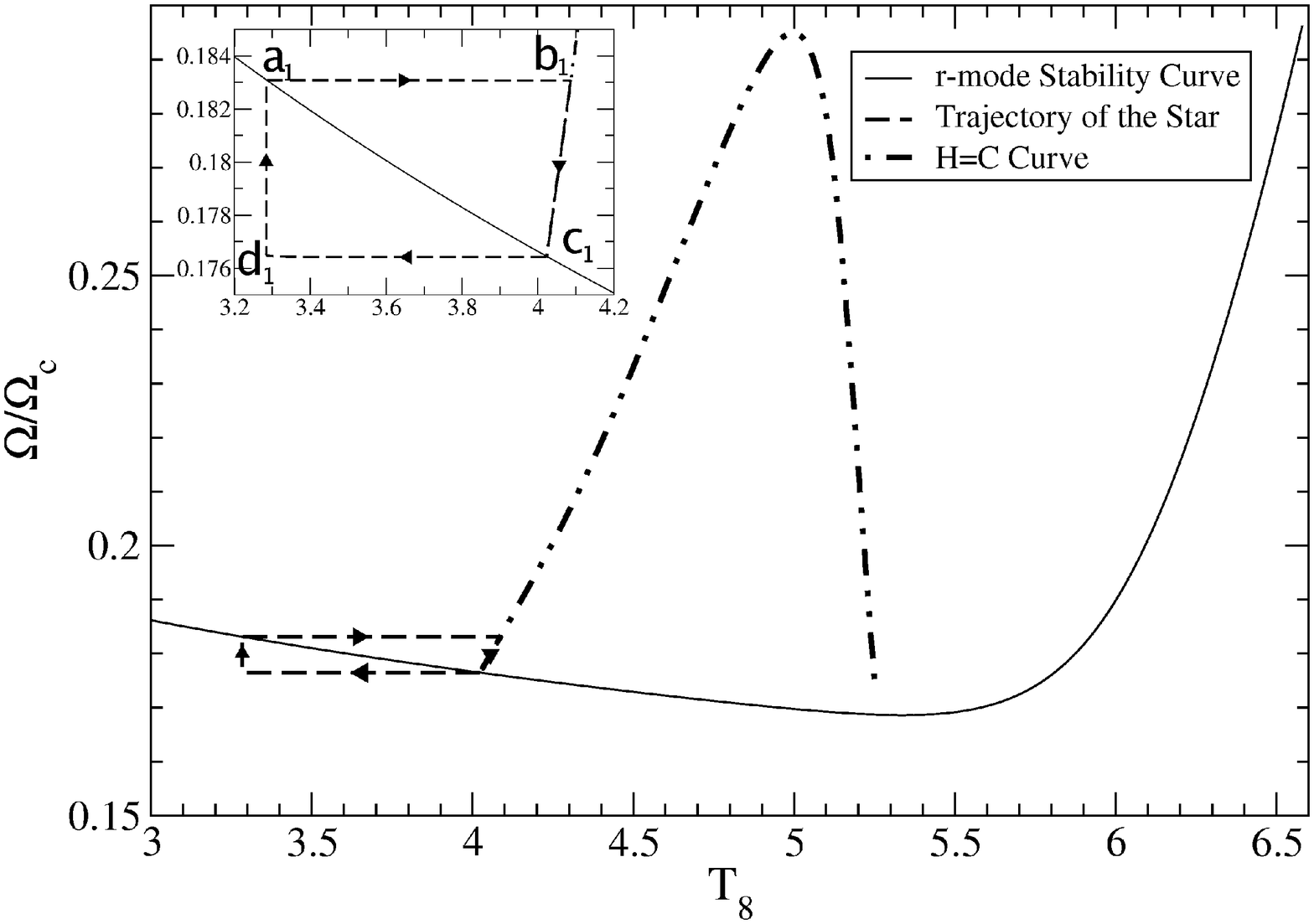}
\epsfxsize=240pt
\epsfbox{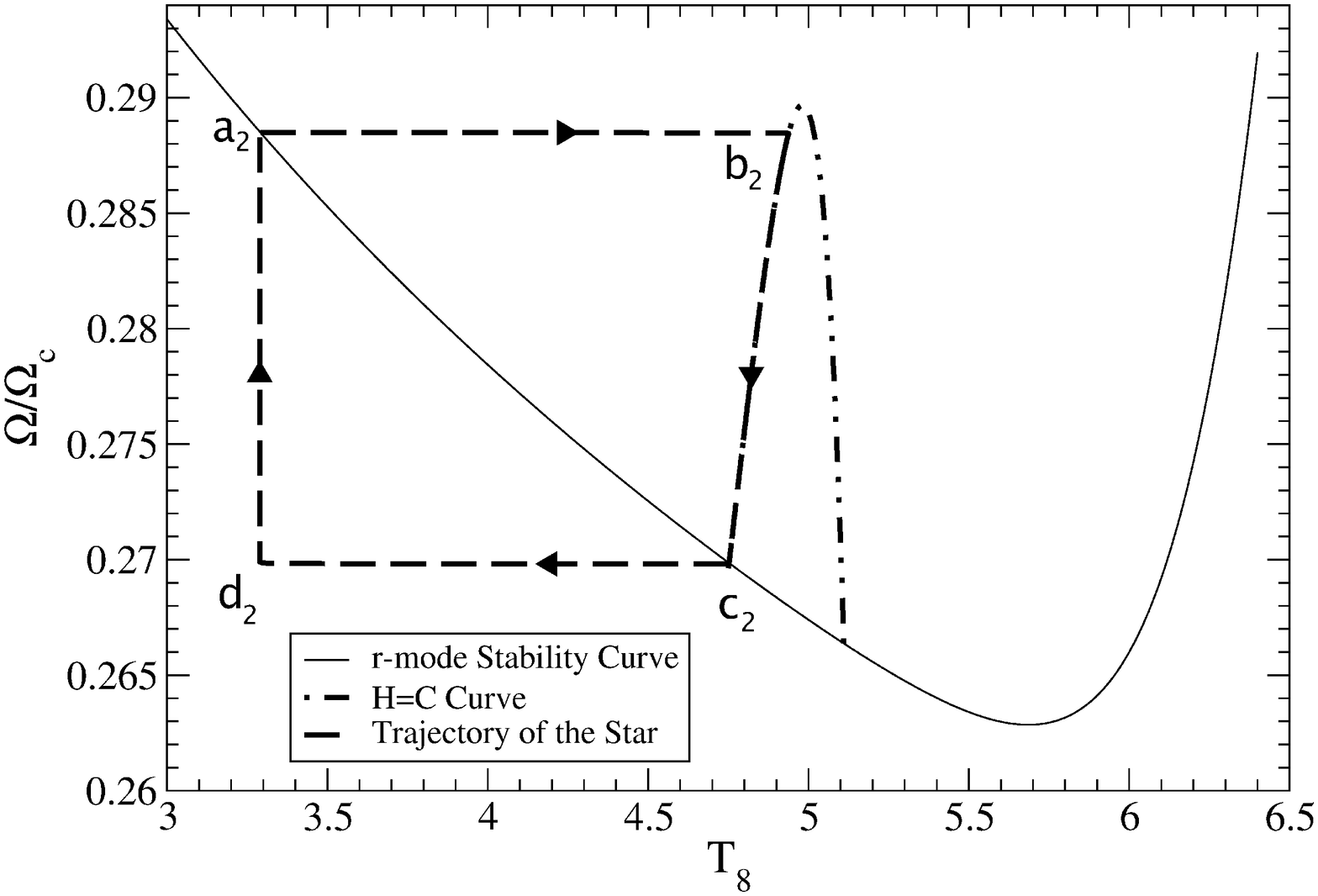}
\caption{Two cyclic trajectories in the $\tilde{\Omega}-T_8$ plane are displayed for a star with $T_c = 5.0\times 10^9$ K  and (a) $f_{\rm{dU}} = 0.15$ and $S_{\rm{ns}}= 0.10$, and (b) $f_{\rm{dU}} = 0.142$ and $S_{\rm{ns}}= 0.35$, which is close to the border between the stable and unstable region (see Fig.\ \ref{scenario}(b)). The thick solid line labeled as the Heating = Cooling ($H=C$) curve is the locus of points in this phase space where the neutrino cooling is equal to the viscous heating due to the unstable modes. The other solid line representing the r-mode stability curve is defined by setting the gravitational driving rate equal to the viscous damping rate. The part of the curve that decreases with $T_8$ is dominated by boundary layer and shear viscosity, while the part of the curve that has a positive slope is dominated by hyperon bulk viscosity. In portion $a_1 \to b_1$ of the trajectory the star heats up at constant $\tilde{\Omega}$. Part $b_1 \to c_1$ represents the spin down stage, 
which occurs when the viscous heating is equal to the neutrino cooling. $c_1 \to d_1$ shows the star cooling back to the initial $T_8$. Segment $d_1 \to a_1$ displays the accretional spin-up of the star back to the r-mode stability curve.  The cycle $a_2 \to d_2$ proceeds in the same way. This cycle is close to the peak of the $H=C$ curve. Configurations above this peak will run away.}
 \renewcommand{\arraystretch}{0.75}
 \renewcommand{\topfraction}{0.6}
\label{cycle}
\end{center}
\end{figure}

 Fig.\ \ref{scenario}(a) shows possible evolutionary trajectories of a neutron star in the angular velocity-temperature $\tilde{\Omega}-T_8$ plane, where $T = T_8 \times 10^8$ K is the core temperature, and $\tilde{\Omega} = \Omega/\Omega_c =\Omega/\sqrt{\pi G \bar{\rho}}$ with $\bar{\rho}$ the mean density of the neutron star. Fig.\ \ref{scenario}(b) displays the regions in $f_{\rm{dU}} - S_{\rm{ns}}$ in which the trajectories occur. Here $f_{\rm{dU}}$ represents the fraction of the star that is above the density threshold for direct URCA reactions and $S_{\rm{ns}}$ is the slippage factor that reduces the relative motion between the crust and the core taking into account the elasticity of the crust \cite{LU}. The stability regions are shown at fixed hyperon superfluidity temperature, $T_c = 5.0 \times 10^9$ K.  The initial part of the evolution is similar in all scenarios and can be  divided into phases. \newline
 \label{SecScenarios}
{\bf Phase 0}. Spin up below the r-mode stability curve at $T_8=T_{8\, \rm{in}}$ such that nuclear heating balances neutrino cooling. \newline
 {\bf Phase 1}.\ Linear regime. The r-mode amplitude grows exponentially. The phase ends when the r-mode reaches the parametric instability. \newline
{\bf Phase 2}.\ The triplet coupling leads to quasi-steady mode amplitudes. The star is secularly heated at approximately constant $\Omega$ because of viscous dissipation in all three modes. \newline
{\bf  Phase 3}.  Several trajectories are possible depending on how the previous phase ends.  \newline
{\bf a.} {\bf Fast Runaway.} The star fails to reach thermal equilibrium when the trajectory passes over the peak of the Heating = Cooling ($H=C$) curve. This leads to rapid runaway. The daughter modes damp eventually as bulk viscosity becomes important, and the r-mode grows exponentially until the trajectory hits the r-mode stability curve again. This  scenario ends as predicted by Nayyar and Owen \cite{mohit}. However, the r-mode passes its second parametric instability threshold soon after it starts growing again. This  requires the inclusion of more modes to follow the evolution, which is the subject of future work. \newline
 {\bf b.} The star reaches thermal equilibrium. There are then three possibilities: \newline
 {\bf (i)} {\bf Cycle}. The star cools and spins down slowly, descending the $H=C$ curve until it crosses the r-mode stability curve again. At this point the instability shuts off. The star cools back to $T_{8\, \rm{in}}$ at constant $\tilde{\Omega}$ and then the cycle repeats itself.  At $T_c = 5.0 \times 10^9$ K this scenario occurs for values of $S_{\rm{ns}} < 0.50$ and large enough values of $f_{\rm{dU}}$. However, if $T_c$ is larger, the cycle region in the $f_{\rm{dU}}$-$S_{\rm{ns}}$ phase space increases dramatically (see Fig.\ \ref{Fig9}(a)). Note that our cycles are different from those obtained by Levin \cite{levin} in that the spin-down phase does not start when the r-mode amplitude saturates (or in our case when it reaches the parametric instability threshold),  but rather when the system reaches thermal equilibrium. The r-mode amplitude does not grow significantly above its first parametric instability threshold, remaining close to $\sim 10^5$ and so the part of the cycle in which the r-mode is unstable also lasts longer than in Ref.\ \cite{levin}. Also, our cycles are narrow.  During spin-down the temperature changes by less than 20 \% and $\tilde{\Omega}$ changes by less than 10\% of the initial value. (See Sec.\ \ref{cycle} for a detailed example.)  \newline
 {\bf (ii)} {\bf  Steady State.} For small $S_{\rm{ns}}$ and large enough $f_{\rm{dU}}$ ($f_{\rm{dU}} \gtrsim 5 \times 10^{-5}$, $S_{\rm{ns}} \lesssim 0.04$; see Fig. \ref{scenario}(b)) the star evolves towards an $\tilde{\Omega}$ equilibrium.  The trajectory either ascends or descends  the $H=C$ curve (spins up and heats or spins down and cools). The evolution stops when the accretion torque equals the gravitational radiation emission. 
 \newline
   {\bf (iii)} {\bf Slow Runaway}. For small $S_{\rm{ns}}$ and very small $f_{\rm{dU}}$ ($S_{\rm{ns}} \lesssim 0.03$, $f_{\rm{dU}} < 5 \times 10^{-5} $) the star ascends the $H=C$ curve until the peak is overcome and subsequently a runaway occurs. The daughter modes eventually damp and the r-mode grows exponentially until it crosses its second parametric instability threshold and more modes need to be included. \newline
   
     Bulk viscosity only affects the runaway evolutions; the cyclic and steady state evolutions found here would be the same if there were no hyperon bulk viscosity.  For large $T_c\sim 10^{10}$, or for models with no hyperons at all, there would be no runaway region (See Fig.\ \ref{Fig9}(a) for an $f_{\rm{dU}}-S_{\rm{ns}}$ scenario space with a larger $T_c = 6.5 \times 10^9$ K where the fast runaway region has shrunk dramatically and the slow runaway region has disappeared.)
     %For $T_c \lesssim 4.5  \times 10^9$ K the stable regions are completely suppressed. At very low critical temperatures $T_c \lesstsim 3 \times 10^9$ K the star enters the regime studied by Ref. \cite{Wagoner} in which the evolutions start on a part of the r-mode stability that has a positive slope and the r-mode amplitude saturates below the lowest parametric instability threshold.    
\section{Possible Evolution Scenarios}
In this section we examine examples of the different types of evolution in more detail. We assume $\dot{M} = 10^{-8} M_\odot$/yr and $T_c = 5.0 \times 10^9$ K. 
\subsection{Cyclic Evolution}
\label{SecCycle}
%(the star spins down and cools down from $T_8=4.10, \; \tilde{\Omega} = 0.1830$ to $T_8=4.02,  \; \tilde{\Omega} = 0.1765$)
In this sub-section we present the features of typical cyclic trajectories of neutron stars in the angular velocity temperature plane in more detail. We focus on two cases: (C1) $S_{\rm{ns}} = 0.10$ and $f_{\rm{dU}} = 0.15$ and (C2) $S_{\rm{ns}} = 0.35$ and $f_{\rm{dU}} = 0.142$. In this scenario the 3-mode system is sufficient to model the nonlinear effects and successfully stops the thermal runaway. The numerical evolution is started once the star reaches the r-mode stability curve. The initial temperature of the star is at the point where nuclear heating equals neutrino cooling in Eq.\ (\ref{thermalevol}) that is approximately $T_{8\,\rm{in}}\approx 3.29$ for both cases. The initial $\Omega$ is the angular velocity that corresponds to this temperature on the r-mode stability curve, which differs for the different $S_{\rm{ns}}$ ($\tilde{\Omega}_{\rm{in}} = 0.183$ for $C_1$ and $\tilde{\Omega}_{\rm{in}} = 0.288$ for $C_2$).

Figs.\  \ref{cycle}(a) and (b) display the cyclic evolution for trajectories $C_1$ and $C_2$ of Fig.\ \ref{scenario}(b). In leg $a_1 \to b_1$ of the trajectory the r-mode and, once the r-mode amplitude increases above the first parametric instability threshold, the two daughter modes it excites, viscously heat up the star until point $b_1$ when the neutrino cooling balances the viscous dissipation. This part of the evolution occurs at constant angular velocity over a period of $t_{\rm{heat-up}} \approx 100 \; \rm{yr}$ and a total temperature change $(\Delta T)_{a_1-b_1} \approx 0.80 $ ($\approx 24$\% of $T_{8\,\rm{in}}$). The points where the viscous heating  compensates the neutrino cooling are represented by the Heating = Cooling ($H=C$) curve. This is determined by setting Eq.\ \ref{thermalevol} to zero and using the quasi-stationary solutions given by Eq.\ (\ref{stationarySol}) for the three modes on the right hand side. The star continues to evolve on the $H=C$ curve for part $b_1 \to c_1$ of the trajectory as it spins down and cools down back to the r-mode stability curve. This spin-down stage lasts a  time $t_{\rm{spin-down} \, b_1 - c_1} \approx 23,000 \; \rm{yr}$ that is much longer than the heat-up period.   This timescale is very sensitive to changes in the slippage factor and can reach $10^6$ yr for smaller values of $S_{\rm{ns}}$ that are close to boundary of the steady state region. The cycle is very narrow in angular velocity with a total angular velocity change of less than 4\%, $(\Delta \tilde{\Omega})_{b_1-c_1}\approx 0.0066$. The temperature also changes by only about 2\%, $(\Delta T_8)_{b_1-c_1} \approx 0.08$ in this spin-down period.  Segment $c_1 \to d_1$ represents the cooling of the star to the initial temperature on a timescale of $\sim 2, 000$ yr. In part $d_1 \to a_1$ the star spins up by accretion at constant temperature back to the original crossing point on the r-mode stability curve. This last part of the trajectory is the longest-lasting one, taking  $ \approx 200, 000$ yr at our chosen $\dot{M}$ of $10^{-8} M_\odot \rm{yr}^{-1}$.  The cycle $C_2$ in Fig.\  \ref{cycle}(b) proceeds in a similar fashion. It is important to note that this configuration is close to the border between the ``FAST RUNAWAY" and ``CYCLE" regions and therefore close to the peak of the $H=C$ curve. Configurations above this peak (e.g., with the same $f_{\rm{dU}}$ and higher $S_{\rm{ns}}$) will go through a fast runaway.

\begin{figure}
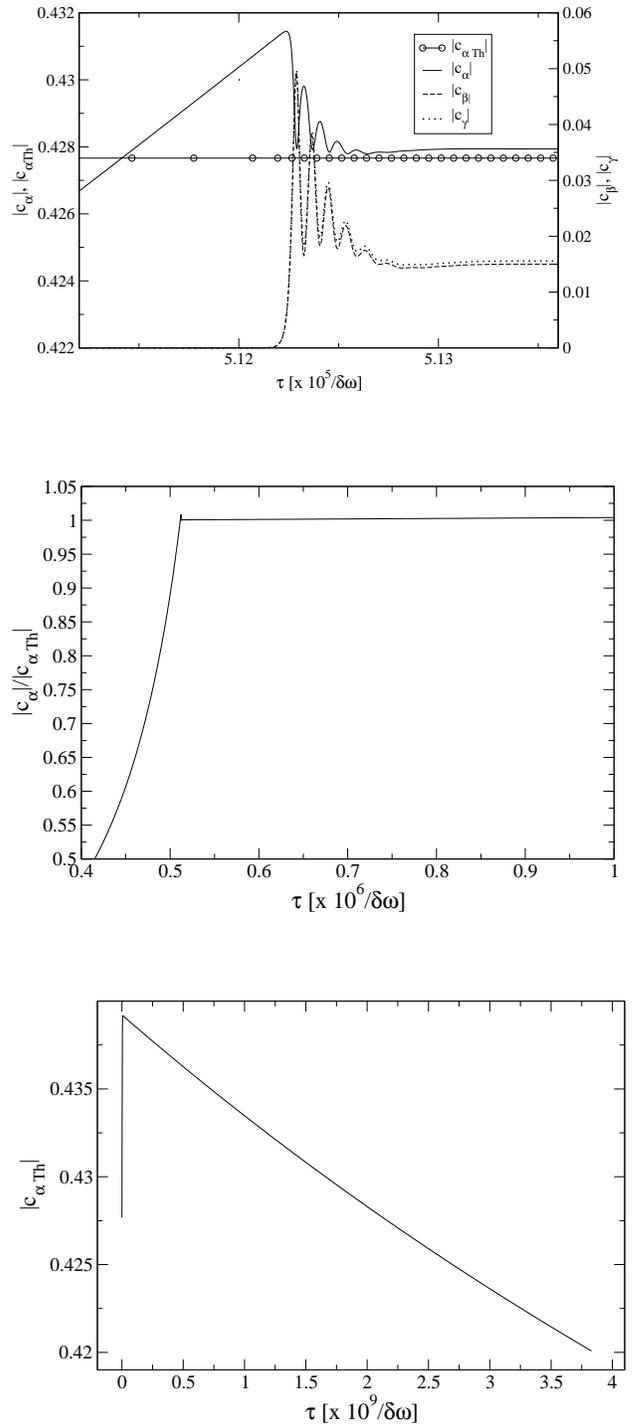

\begin{center}
\epsfxsize=230pt
\epsfbox{Fig3a.eps}
\newline
\newline
\newline
\newline
\epsfxsize=230pt
\epsfbox{Fig3b.eps}
\newline
\newline
\newline
\newline
\epsfxsize=230pt
\epsfbox{Fig3c.eps}
% \epsfysize=0pt
% \epsfbox{Fig3b.eps}
% \epsfysize=0pt
% \epsfbox{Fig3c.eps}
\caption{(a)The amplitudes of the r-mode $|C_\alpha|$ and of the $n=13, m=-3$ and $n=14, m=1$ inertial modes $|C_\beta|$ and $|C_\gamma|$ are shown as a function of time for a star that executes a cyclic evolution (same parameters as in Fig.\ \ref{cycle}). The lowest parametric instability threshold is also displayed.  (b)The ratio of the r-mode amplitude to the parametric instability threshold is plotted as a function of time. It can be seen that once the r-mode crosses the parametric instability threshold it remains close to it for the rest of the evolution. (c)The parametric instability threshold is displayed as a function of time. Its value changes as the angular velocity and temperature evolve.}
\label{amplitudes}
\end{center}
\end{figure}

Fig.\ \ref{amplitudes}(a) shows the evolution of the three modes in the first few years after the star first reaches the r-mode stability curve. In this region the r-mode is unstable and initially grows exponentially. Once it has increased above the first parametric instability threshold the daughter modes are excited.  The oscillations of the three modes display some of the typical dynamics of a driven three-mode system. When the r-mode transfers energy to the daughter modes they increase exponentially while the r-mode decreases. Similarly, when daughter modes decrease the r-mode increases.  The viscosity damps the oscillations and the r-mode amplitude settles at a value close to the parametric instability threshold. Fig.\ \ref{amplitudes}(b) displays the evolution of the r-mode amplitude divided by the parametric instability threshold on a longer timescale. It can be seen that the r-mode never grows significantly beyond this first threshold. Fig.\ \ref{amplitudes}(c) shows the evolution of the parametric instability threshold as a function of time. The threshold  increases as the temperature increases and the star is viscously heated by the three modes. When the star spins down in thermal equilibrium, the threshold decreases to a value close to its initial value.  
\subsection{Steady State Evolution}
\label{SecSteady}
\begin{figure}
\begin{center}
\leavevmode
\epsfxsize=250pt
\epsfbox{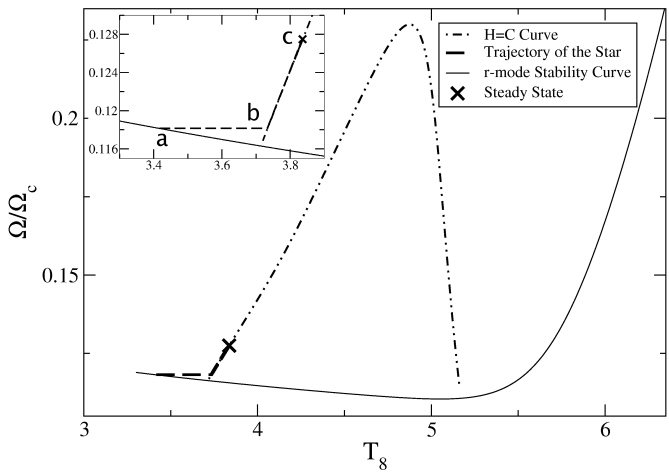}
\caption{The trajectory of a neutron star in the $\tilde{\Omega}-T_8$ phase space is shown for a model with $T_c = 5.0 \times 10^9$ K, $f_{\rm{dU}} = 0.03$ and $S_{\rm{ns}} = 0.03$ that reaches an equilibrium steady state.  The star spins up until it crosses the r-mode stability curve and the r-mode becomes unstable.  The r-mode then quickly grows to the first parametric instability threshold and excites the daughter modes. In leg $a \to b$ of the trajectory the star is viscously heated by the mode triplet until the system reaches thermal equilibrium. Segment $b \to c$ shows the star continuing to heat and spin up in thermal equilibrium until the accretion torque is balanced by the gravitational radiation emission. The r-mode stability curve represents the points in phase space where the viscous driving rate is equal to the gravitational driving rate. The H=C curve is the locus of points where the viscous dissipation due to the mode triplet balances the neutrino cooling.}
 \renewcommand{\arraystretch}{0.75}
 \renewcommand{\topfraction}{0.6}
\label{SteadyStateOmT}
\end{center}
\end{figure}

\begin{figure}
\begin{center}
\leavevmode
\epsfxsize=250pt
\epsfbox{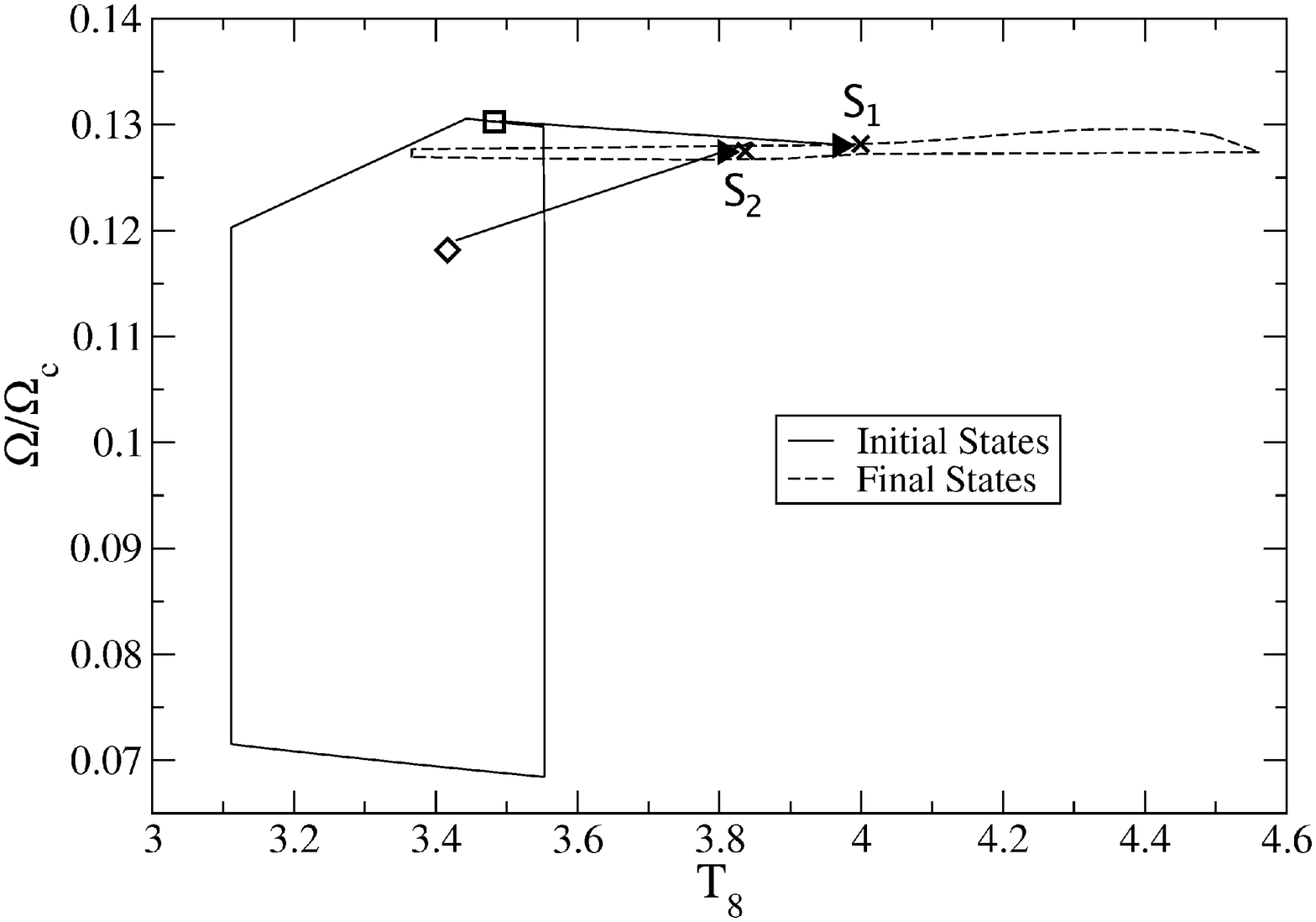}
\caption{The ($\tilde{\Omega}$, $T_8$) initial values (region delimited by the solid line) that lead to equilibrium steady states and their corresponding final steady state values (region enclosed by the dashed line) are shown.  Since both the initial and final values of $T_8$ are low, these evolutions are roughly independent of $T_c$.
}
 \renewcommand{\arraystretch}{0.75}
 \renewcommand{\topfraction}{0.6}
\label{SteadyStatePhaseSpace}
\end{center}
\end{figure}
This sub-section focuses on evolutions that lead to a steady equilibrium state in which the rate of accretion of angular momentum is balanced by the rate of loss via gravitational radiation emission.  This scenario is restricted to stars with small slippage factor ($S_{\rm{ns}} \lesssim 0.04$, see Fig. \ref{scenario}(b)) and boundary layer viscosity.  A typical trajectory of a star that reaches such an equilibrium is shown in Fig. \ref{SteadyStateOmT}.  As always, we start the evolution at the point on the r-mode stability curve at which the nuclear heating balances neutrino cooling. Above the r-mode stability curve the gravitational driving rate is greater than the viscous damping rate
and the r-mode grows exponentially until nonlinear effects become important. In this case, as in the cyclic evolution, the triplet of modes at the lowest parametric instability threshold is sufficient to stop the thermal runaway. The r-mode remains close to the first instability threshold for the length of the evolution and after a few oscillations the three modes settle into their quasi-stationary states, which change only secularly as the spin and temperature of the star evolve. The modes heat the star viscously at constant $\tilde{\Omega}$ in segment $a \to b$ of the trajectory for $t_{\rm{heat-up}} \approx 1, 100 \; \rm{yr}$. At point $b$, the star reaches a state of thermal balance. In leg $b \to c$ the star continues its evolution in thermal equilibrium and slowly spins up due to accretion until the angular velocity evolution also reaches an equilibrium. The timescale to reach an equilibrium steady state is $t_{\rm{steady}} \approx 3.5 \times 10^6\;\rm{yr}$ for this set of parameters.

Fig. \ref{SteadyStatePhaseSpace} displays the possible initial values for the angular velocity $\tilde{\Omega}$ and temperature $T_8$ of the star that lead to a balancing between the accreted angular momentum and the angular momentum emitted in gravitational waves. The fraction of the star that is above the threshold for direct URCA reactions and the slippage factor are varied within the corresponding ``STEADY STATE" region of Fig. \ref{scenario}(b). The final equilibrium values are also displayed and cluster in a narrower region than the initial values. Because viscosity is so small in this regime, the values of $\Omega$ also tend to be small. Thus, although an interesting physical regime,  this case is most likely not relevant to recycling by accretion to create pulsars with spin frequencies as large as 716 Hz. Note that a steady state can be achieved when $S_{\rm{ns}} = 0$. This is the probable end state of the problem first calculated by Levin \cite{levin}. The reason we do not find a cycle at low $S_{\rm{ns}}$ is twofold: (1) the shear viscosity we are using is lower (shear viscosity in Ref. \cite{levin} is amplified by a factor of 244), and (2) the nonlinear couplings keep all mode amplitudes small. 
\subsection{Thermal Runaway Evolutions}
\label{SecRun}
We now consider evolutions in which the three-mode system is not sufficient to halt the thermal runaway. We observe two such scenarios. In the first scenario, the star is unable to reach thermal equilibrium. The runaway occurs on a period much shorter than the accretion timescale and so the whole evolution is at approximately constant angular frequency. In the second scenario, the star reaches a state of thermal equilibrium but the spin evolution does not reach a steady state. The star continues to spin up by accretion until it climbs to the peak of the $H=C$ curve, thermal equilibrium fails and a runaway occurs. 
\subsubsection{Fast Runaway}

\begin{figure}
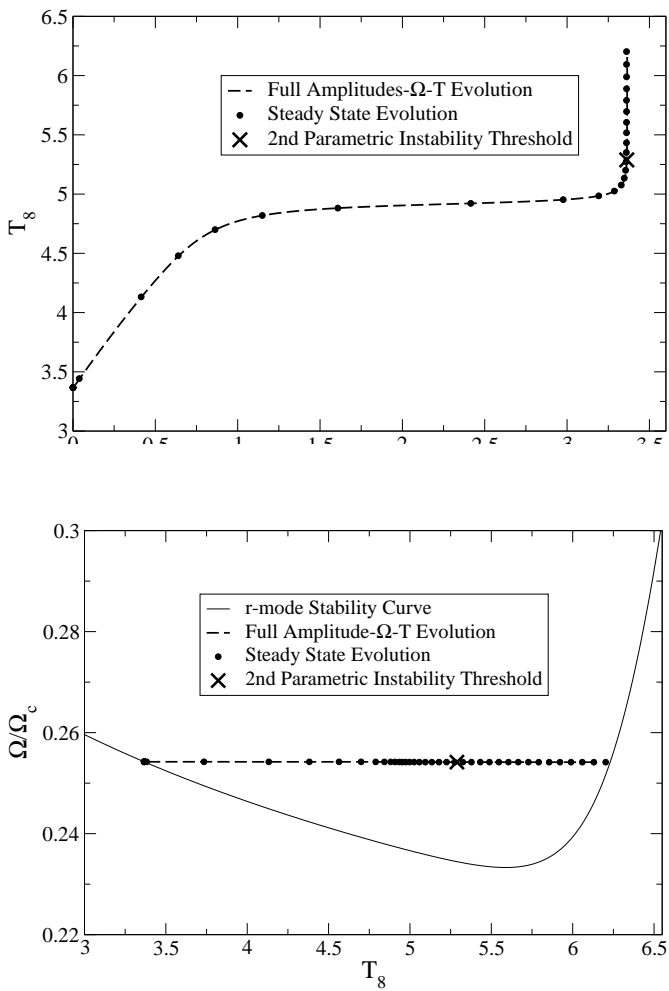

\begin{center}
\leavevmode
\epsfxsize=250pt
\epsfbox{Fig6a.eps}
\epsfxsize=250pt
\newline\newline
\epsfbox{Fig6b.eps}
\caption{This plot compares the full evolution resulting from solving Eqs. (\ref{eqcode}),(\ref{ang2}),(\ref{thermalevol}) with the reduced $\Omega-T$ evolution that assumes the amplitudes go through a series of steady states Eqs. (\ref{spinSteady})-(\ref{thermalSteady}) for a model with $T_c = 5.0 \times 10^9$ K, $f_{\rm{dU}} = 0.058$ and $S_{\rm{ns}} = 0.25$.
(a) The temperature is displayed as a function of time for the two different methods. (b) The angular velocity $\tilde{\Omega} = \Omega/\Omega_c$ is shown as a function of temperature. The evolution occurs at constant spin frequency. It can be seen that the steady-state amplitude approximation is extremely good. The `X' shows the point at which the r-mode crosses its second lowest parametric instability threshold, where additional dissipation would become operative. }
 \renewcommand{\arraystretch}{0.75}
 \renewcommand{\topfraction}{0.6}
\label{SteadyvsFull}
\end{center}
\end{figure}

A typical trajectory of a star that goes through a rapid thermal runaway is displayed in Fig.\ \ref{SteadyvsFull}. This star has $S_{\rm{ns}} = 0.25$ and $f_{\rm{dU}} = 0.058$. Initially, the growth of the r-mode is halted by the two daughter modes once the lowest parametric instability threshold is crossed, and  the three modes settle in the ($\Omega$,$T$)-dependent quasi-stationary states of Eqs.\ (\ref{stationarySol}).  They viscously heat up the star until hyperon bulk viscosity becomes important for the daughter modes.
As the amplitudes of the daughter modes decrease the coupling is no longer strong enough to drain enough energy to stop the growth of the r-mode.  The daughter modes are completely damped and the r-mode increases exponentially. The system goes back to the one-mode evolution described by Ref.\ \cite{mohit}.

Fig.\ \ref{SteadyvsFull}(a) and (b) compare both  the temperature evolution and the trajectory in the $\tilde{\Omega}-T_8$ plane of the star for a simulation solving the full set of equations to a simulation that assumes quasi-stationary solutions for the three amplitudes and evolves only the angular velocity and temperature of the star. It can be seen that the steady state approximation is very good until the thermal runaway occurs. Afterward, the temperature evolution of the reduced equations is offset slightly  from the quasi-steady result and intersects the r-mode instability curve sooner. This evolution is similar to that described by Nayyar and Owen \cite{mohit}. However, the r-mode crosses its second lowest parametric instability much earlier in the evolution (see the `X' in the figure), and at that point more modes need to be included to model the instability accurately. Thus, we cannot be sure  that a runaway must occur in this case. We shall return to this issue in a subsequent paper.

\subsubsection{Slow Runaway}
\begin{figure}
\begin{center}
\leavevmode
\epsfxsize=250pt
\epsfbox{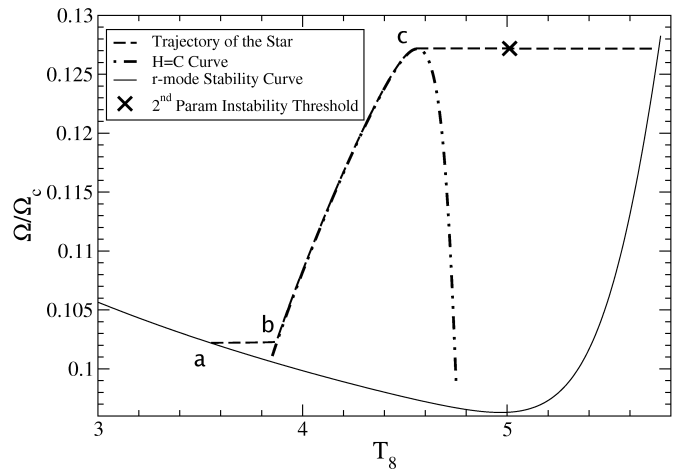}
\caption{The trajectory of a neutron star in the $\tilde{\Omega}-T_8$ phase space is shown for a model with $T_c = 5.0 \times 10^9$ K, $f_{\rm{dU}} = 4.0 \times 10^{-5}$ and $S_{\rm{ns}} = 0.02$ that goes through a slow thermogravitational runaway. Portion $a \to b$ of the trajectory shows the mode triplet heating up the neutron star through boundary layer and shear viscosity until the system reaches  thermal equilibrium. Segment $b \to c$ represents the accretional spin-up of the star in thermal equilibrium. The dotted-dashed line is the locus of points where the viscous dissipation of the mode triplet is equal to the neutrino cooling, and is labeled as the $H=C$ curve. The star reaches the maximum of this curve and fails to reach an equilibrium between the accretion torque and gravitational emission. It then continues heating at constant angular velocity and crosses its second lowest parametric instability threshold, at which point more modes would need to be included to make the evolution accurate. Eventually the star reaches the r-mode stability curve again.}
 \renewcommand{\arraystretch}{0.75}
 \renewcommand{\topfraction}{0.6}
\label{slowrun}
\end{center}
\end{figure}
In this section we examine evolutions in which the neutron star has both a very small slippage factor, $S_{\rm{ns}} \lesssim 0.03$, and only a small percentage of the star is above the threshold for direct URCA reactions, $f_{\rm{dU}} < 5 \times 10^{-5}$. A trajectory for this kind of evolution is displayed in Fig. \ref{slowrun}. After the star crosses the r-mode stability curve, the r-mode increases beyond the first parametric instability threshold, and its growth is temporarily stopped by energy transfer to the daughter modes. As in the previous scenarios, the star is viscously heated by the mode triplet at constant $\Omega$ in part $a \to b$ of the trajectory on a timescale of about $5, 000$ yr. At point $b$, it reaches thermal equilibrium. In leg $b \to c$ of the trajectory, the star continues its evolution by ascending the $H=C$ curve and spinning up because of accretion for about $2 \times 10^6$ yr without finding an equilibrium state for the angular momentum evolution. Once it reaches the peak of the $H=C$ curve, the cooling is no longer sufficient to stop the temperature from increasing exponentially and a thermal runaway occurs. The cross mark `X' on the trajectory shows the point at which the r-mode amplitude crosses its second lowest parametric instability threshold. At this stage more inertial modes  need to be included to model the rest of this evolution correctly. As for the cases that evolve to  steady states, these long-timescale runaways tend to occur at low spin rates. 
%\subsubsection{Stability analysis: Expansion around the steady state}
%--analytic estimate for the timescale to reach the steady state
\section{Probability of Detection}
\label{detection}
\begin{figure}
\begin{center}
\leavevmode
\epsfxsize=250pt
\epsfbox{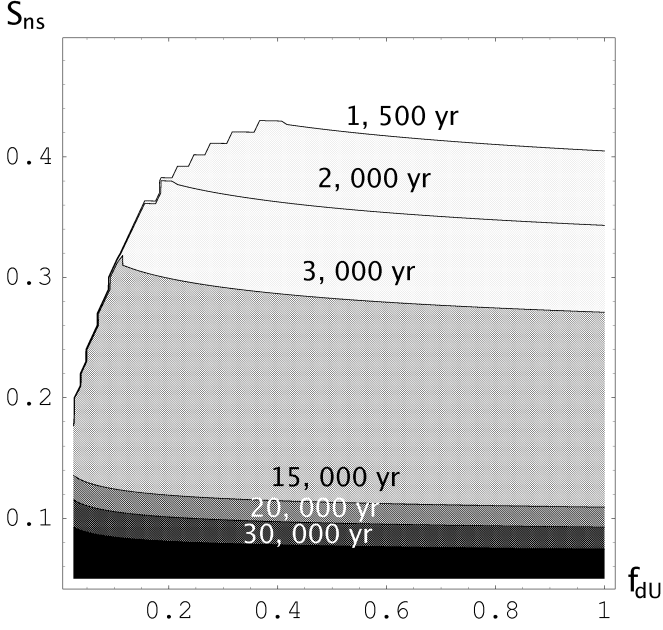}
\caption{The spin-down timescale is shown as slippage factor $S_{\rm{ns}}$ and fraction of the star subject to direct URCA $f_{\rm{dU}}$ for cyclic evolutions are varied for a fixed hyperon critical temperature of $T_c = 5.0 \times 10^9$ K. This timescale dominates  the heat-up timescale and hence represents the time the star spends above the r-mode instability curve. It increases as the viscosity is lowered and the star gets closer to the steady state region. 
}
 \renewcommand{\arraystretch}{0.75}
 \renewcommand{\topfraction}{0.6}
\label{Timescales}
\end{center}
\end{figure}
\begin{figure}
\begin{center}
\leavevmode
\epsfxsize=250pt
\epsfbox{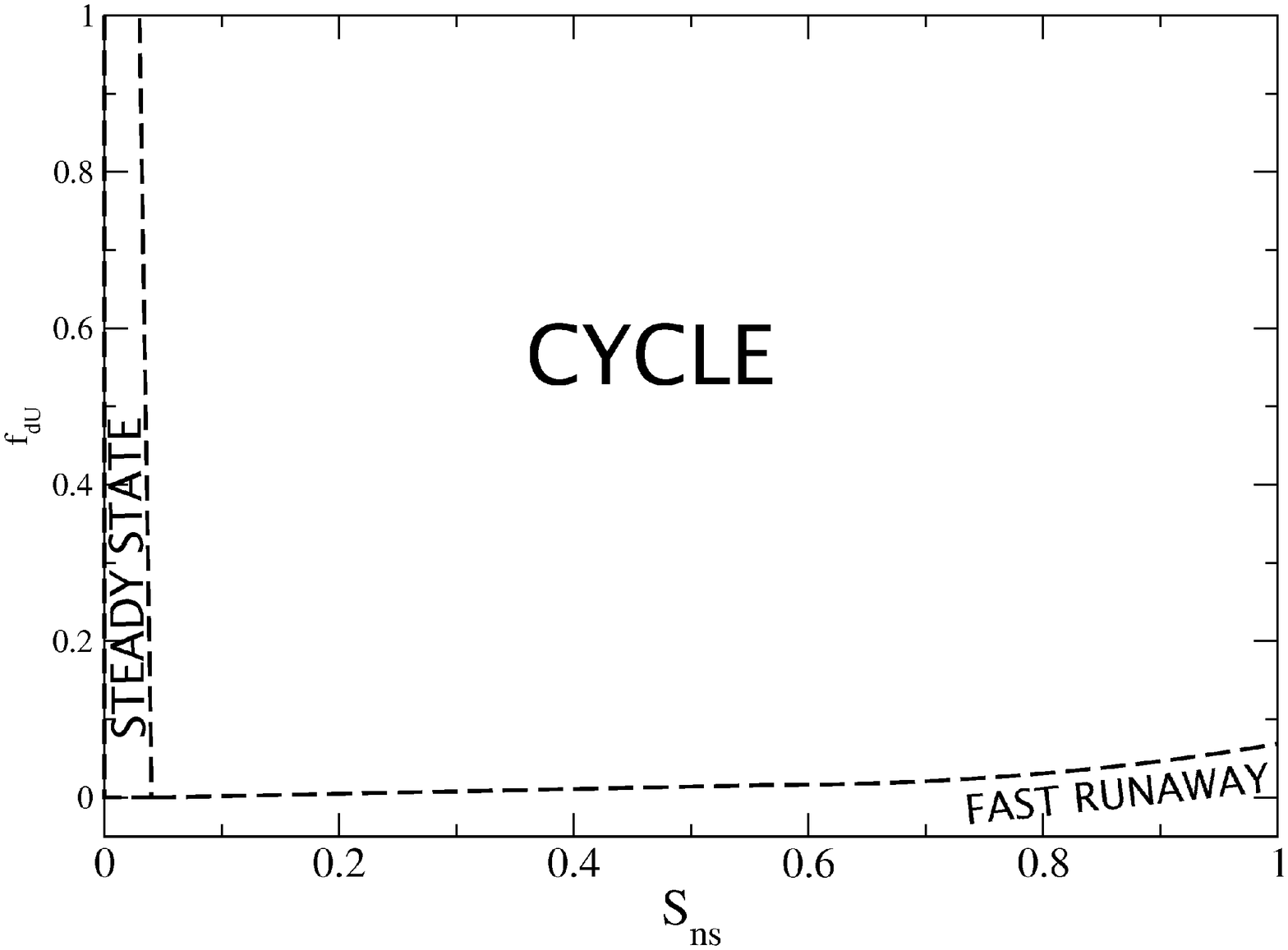}
\epsfxsize=250pt
\epsfbox{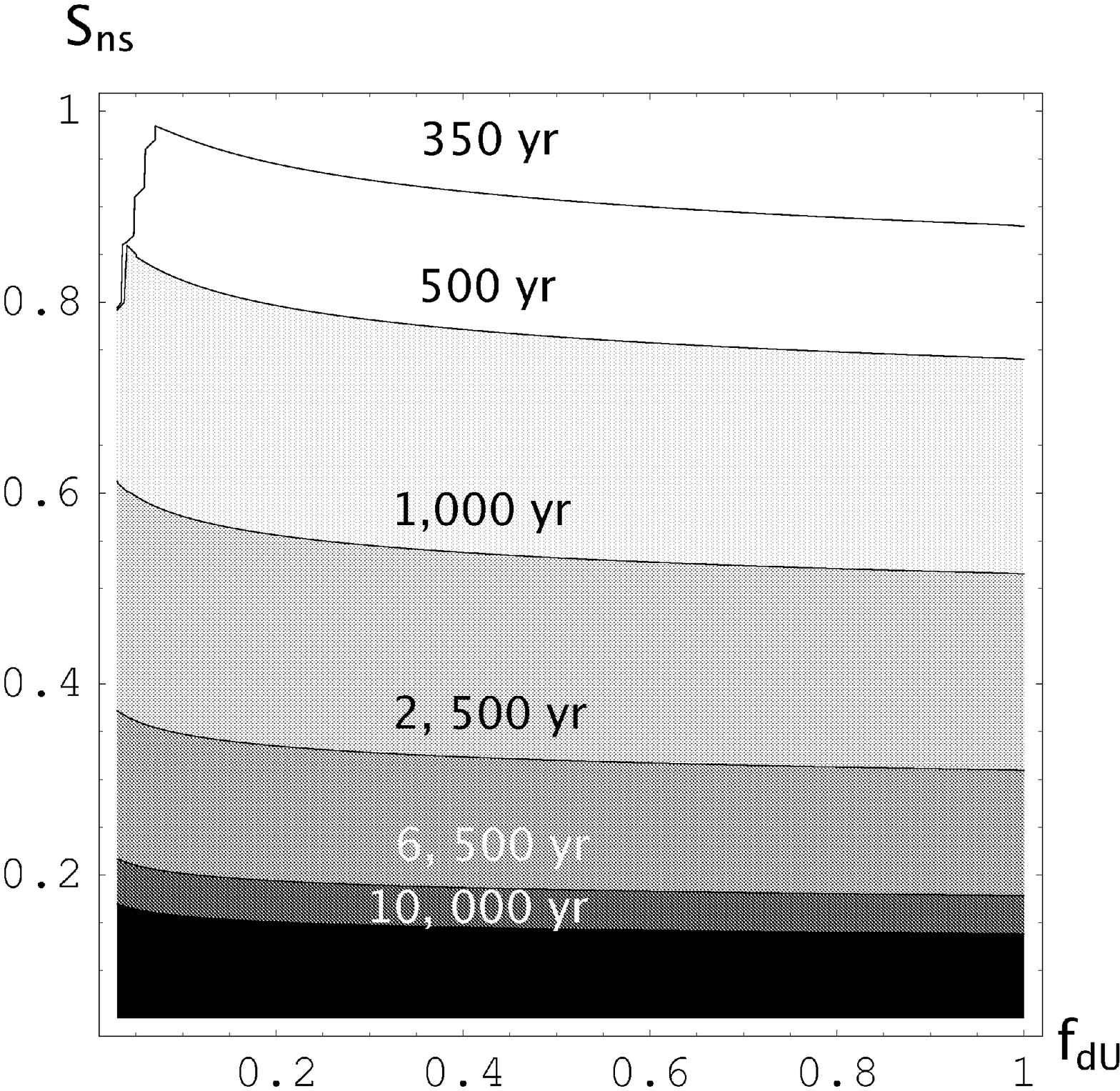}
\caption{
(a)The stability regions are plotted at fixed hyperon superfluidity temperature $T_c = 6.5 \times 10^9$ K, while varying  $f_{\rm{dU}}$ and $S_{\rm{ns}}$. The steady state region remains roughly the same as in Fig.\ \ref{scenario}(b), the slow run-away region disappears, and the cycle region increases dramatically while shrinking the fast-runaway region. (b) The spin-down timescale is shown for the cyclic evolutions in part (a). }
 \renewcommand{\arraystretch}{0.75}
 \renewcommand{\topfraction}{0.6}
\label{Fig9}
\end{center}
\end{figure}
Fig.\ \ref{Timescales} shows how the time the star spends above the r-mode stability curve changes when $S_{\rm{ns}}$ and $f_{\rm{dU}}$ are varied. For large enough values of $S_{\rm{ns}}$ the boundary layer viscosity dominates. In this region of phase space the spin-down timescale can be approximated by
\begin{eqnarray}
\label{SpinDown}
t_{\rm{spin-down}} &=& \int_b^c \frac{dt}{d\tilde{\Omega}} d \tilde{\Omega}  \\ \nonumber
&\approx& \frac{\tilde{I} \tau^0_{\rm{GR}}}{6} \frac{(4 \tilde{\kappa})^2 \tilde{\omega}_\beta \tilde{\omega}_\gamma}{|\delta \tilde{\omega}|^2} \frac{1}{|\bar{C}_\alpha|^2} \frac{\Delta \tilde{\Omega}}{\rm{<}\tilde{\Omega}\rm{>}^6}  \\ \nonumber
&\approx& 250\,\rm{yr} \, \frac{\Delta \nu_{\rm{kHz}}}{\rm{<}\nu_{\rm{kHz}}\rm{>}^7} \frac{1}{M_{1.4} R_6^4} \left(\frac{|c^{\rm{th}}_\alpha|}{|c_\alpha|}\right)^2,
\end{eqnarray}
where $M_{1.4} = M/(1.4 M_\odot)$, $R_6 = R/(10^6$cm), $\nu_{\rm{kHz}} = \nu/1 kHz$, $\tilde{I} = 0.261$ \cite{Owen1998}, the r-mode amplitude at its parametric instability threshold $|c_\alpha^{\rm{th}}| \approx |\delta \tilde{\omega}|/(4 \tilde{\kappa}\sqrt{\tilde{\omega}_\beta \tilde{\omega}_\gamma}) \approx 1.5 \times 10^{-5}$, and $\bar{C}_\alpha = \sqrt{\tilde{\Omega}} |c_\alpha|/|c_\alpha|^{\rm{th}}$. This approximation agrees with spin-up timescales obtained from our simulations to $\sim 25\%$.

 The maximum $\nu$ is approximately the same as the initial frequency, and can be determined by equating the driving and damping rate of the r-mode, since it is on the r-mode stability curve
\begin{equation}
\label{freq}
\nu_{\rm{max}} \approx 800 Hz \left(\frac{S_{\rm{ns}}}{M_{1.4} R_6}\right)^{4/11} \frac{1}{T_8^{2/11}}.
\end{equation}
Thus, the spin-down timescale is very sensitive to the slippage factor $t_{\rm{spin-down}} \propto S_{\rm{ns}}^{-24/11} (\Delta \nu_{\rm{kHz}}/\nu_{\rm{kHz}})$. The dependences on $f_{\rm{dU}}$ and accretion rate $\dot{M}$ are much weaker; a rough approximation, obtained by matching direct URCA cooling and nuclear heating, is 
$T_{8\,\rm{in}}$ $\propto$ $\dot{M}^{1/6} f_{\rm{dU}}^{-1/6} R_6^{-1/6} M_{1.4}^{-1/9}$, and
$\nu_{\rm{max}} \propto S_{\rm{ns}}^{4/11} f_{\rm{dU}}^{1/33} \dot{M}^{-1/33} R_6^{-1/3} M_{1.4}^{-34/99}$.
% and $t_{\rm{spin-down}} \propto \dot{M}^{2/11} f_{\rm{dU}}^{-2/11}$. 
The gravitational wave amplitude measured at distance d \cite{brady,Owen2001} is
\begin{eqnarray}
 \label{GWamp}
h &\approx& 1.6 \frac{R}{d} \sqrt{\frac{GM}{\tau^0_{\rm{GR}} c^3}} \tilde{\Omega}^3 |c_\alpha| \\ \nonumber
 &\approx& 3 \times 10^{-25} \left(\frac{10\rm{kpc}}{d}\right) M_{1.4} R_6^3 \nu_{\rm{kHz}}^3 \left(\frac{|c_\alpha|}{c_\alpha^{\rm{th}}}\right).
\end{eqnarray}
Taking $\nu \approx \nu_{\rm{max}}$ gives 
\begin{equation}
h \propto S_{\rm{ns}}^{12/11} M_{1.4}^{-1/33} R_6^2 f_{\rm{dU}}^{1/11} \dot{M}^{-1/11}.
\end{equation}
The maximum distance at which sources could be detected by Advanced LIGO interferometers, assuming $h_{\rm{min}} = 10^{-27}$, \cite{brady} is
\begin{eqnarray}
\label{dist}
d_{\rm{max}} &\approx& 3 \, \rm{Mpc} \, \left(\frac{10^{-27}}{h_{\rm{min}}}\right) M_{1.4} R_6^3 \nu_{\rm{kHz}}^3 \left(\frac{|c_\alpha|}{|c_\alpha^{\rm{th}}|}\right) \\ \nonumber
&\approx&1.5 \, \rm{Mpc} \, \left(\frac{10^{-27}}{h_{\rm{min}}}\right) S_{\rm{ns}}^{\rm{12/11}} M_{1.4}^{-1/11} R_6^{21/11} \\ \nonumber
&\times& T_8^{-6/11} \left(\frac{|c_\alpha|}{|c_\alpha^{\rm{th}}|}\right).
\end{eqnarray}

Eqs.\ (\ref{freq}) and (\ref{dist}) imply that gravitational radiation from the r-mode instability may only be detectable for sources in the Local Group of galaxies. Eq.\ (\ref{freq})  implies that for accretion to be able to spin up neutron stars to $\nu \gtrsim 700$ Hz, we must require $(S_{\rm{ns}}/M_{1.4} R_6 \sqrt{T_{8\rm{in}}})^{4/11} \gtrsim 1$. Assuming this to be true, $d_{\rm{max}} \lesssim 1$-1.5 Mpc.  However, $t_{\rm{spin-down}} \approx 1 000$ yr at most, making detection unlikely for any given source. Moreover, unless $S_{\rm{ns}}$ can differ substantially from one neutron star to another, only those with $\nu$ given by Eq.\ (\ref{freq}) can be r-mode unstable. Slower rotators, including almost all LMXBs, are still in their stable spin-up phases. 

Still more seriously, Fig.\ \ref{scenario}(b) shows that spin cycles are only possible for $S_{\rm{ns}} \lesssim 0.50$, assuming $T_c \approx 5.0 \times 10^9$ K; Eq.\ (\ref{freq}) then implies $\nu \lesssim 450$ Hz. This would restrict detectable gravitational radiation to galactic sources, although the duration of the unstable phase could be longer. 

Within the context of our three mode calculation, $S_{\rm{ns}} > 0.50$, which is needed for explaining the fastest pulsars, would imply fast runaway. There are two possible resolutions to this problem. One is that including additional modes prevents the runaway; we shall investigate this in subsequent papers. The second is that $T_c$ is larger, or that neutron stars do not contain hyperons (e.g., because they are insufficiently dense). Fig.\ \ref{Fig9}(a) shows the same phase plane as Fig.\ \ref{scenario}(b) but with $T_c = 6.5 \times 10^9$ K, and Fig.\ \ref{Fig9}(b) shows the results for $t_{\rm{spin-down}}$ analogous to Fig.\ \ref{Timescales}. Larger $T_c$ permits spin cycles for higher values of $S_{\rm{ns}}$ (and hence $\nu$), but the time spent in the unstable regime is shorter.
\section{Conclusions}
\label{conclusion}
In this paper, we model the nonlinear saturation of unstable r-modes of accreting neutron stars using the triplet of modes formed from the $n=3, m=2$ r-mode and the the first two near resonant modes that  become unstable ($n=13, m=-3$ and $n=14, m=1$) by coupling to the r-mode. This is the first treatment of the spin and thermal evolution including the nonlinear saturation of the r-mode instability to provide a physical cutoff by energy transfer to other modes in the system. The model includes neutrino cooling and shear, boundary layer and hyperon bulk viscosity. We allow for some uncertainties in neutron star physics that is not yet understood by varying the superfluid transition temperature, the slippage factor that regulates the boundary layer viscosity, and the fraction of the star that is above the density threshold for direct URCA reactions.
%{\bf need to say that there enough uncertainties elsewhere; we expect the qualitative behavior discussed here to still be valid. However, we expect that the simple model discussed here gives a good qualitative understanding of the nonlinear effects and identifies the key points. }
 In all our evolutions we find that the mode amplitudes quickly settle into a series of quasi-stationary states that can be calculated algebraically, and depend weakly on angular velocity and temperature. The evolution continues along these sequences of quasi-steady states as long as the r-mode is in the unstable regime.  The spin and temperature of the neutron star can follow several possible trajectories depending on interior physics. The first part of the evolution is the same for all types of trajectories: the star viscously heats up at constant angular velocity.  

If thermal equilibrium is reached, we find several possible scenarios. The star may follow a cyclic evolution, and spin down and cool in thermal equilibrium until the r-mode enters the stable regime. It subsequently cools at constant $\Omega$ until it reaches the initial temperature. At this point the star starts spinning up by accretion until the r-mode becomes unstable again and the cycle is repeated. The time the star spends in the unstable regime is found to vary between a few hundred years (large $S_{\rm{ns}} \sim 1$) and $10^6$ yr (small $S_{\rm{ns}} \sim 0.05$). Our cycles are different from those previously found by Ref.\ \cite{levin} in that our amplitudes remain small, $\sim 10^{-5}$, which slows the viscous heating and causes the star to spend more time in the regime where the r-mode instability is active. Furthermore, we find that the star stops heating when it reaches thermal equilibrium and not when the r-mode reaches a maximum value. The cycles we find are narrow with the spin frequency of the star changing less than 10\% even in the case of high spin rates $\sim 750$ Hz. Other possible trajectories are an evolution toward a full steady state in which the accretion torque balances the gravitational radiation emission, and a very slow thermogravitational runaway on a timescale of $\sim 10^6$ yr. These scenarios occur for very low viscosity ($S_{\rm{ns}} \lesssim 0.04$). Although theoretically interesting, they do not allow for very fast rotators of $\sim 700$ Hz.

Alternatively, if the star does not reach thermal equilibrium, we find that it continues heating up at constant spin frequency until it enters a regime in which the r-mode is no longer unstable. This evolution is similar to that predicted by Nayyar and Owen \cite{mohit}. However, the r-mode grows above its second parametric instability threshold fairly early in its evolution and at this point more inertial modes should be excited and the three-mode model becomes insufficient. Modeling this scenario accurately is subject of future work.

We have focused on cases with $T_c \gtrsim 5 \times 10^9$ K. These are cases for which the nonlinear effects are substantial. In this regime, hyperon bulk viscosity is not important except for thermal runaways where we expect other mode couplings, ignored here, to play important roles. Fast rotation requires large dissipation, as has long been recognized \cite{BU,levin} and these models can only achieve $\nu \gtrsim 700$ Hz if boundary layer viscosity is very large. Alternatively, at lower $T_c \lesssim 3 \times 10^9$ K, large rotation rates can be achieved at r-mode amplitudes below the first parametric instability threshold \cite{Wagoner}. Nayyar and Owen found that increasing the mass of the star for the same equation of state makes the hyperon bulk viscosity become important at lower temperatures \cite{mohit}. Conceivably,  there are accreting neutron stars with relatively low masses that have lower central densities and small hyperon populations. These could evolve as detailed here and only spin up to modest frequencies. Hyperons could be more important in more massive neutron stars leading to larger spin rates and very small steady state r-mode amplitude as found by Wagoner \cite{Wagoner}.

Our models imply small r-mode amplitudes of $\sim 10^{-5}$ and therefore gravitational radiation  detectable by advanced LIGO interferometers only in the local group of galaxies up to a distance of a few Mpc. The r-mode instability puts a fairly stringent limit on the spin frequencies of accreting neutron stars of $\nu_{\rm{max}} \approx 800 Hz [S_{\rm{ns}}/(M_{1.4} R_6)]^{4/11} T_8^{-2/11}$. In order to allow for fast rotators of $\gtrsim 700$ Hz in our models a large boundary layer viscosity
with $(S_{\rm{ns}}/M_{1.4} R_6 \sqrt{T_{8\rm{in}}})^{4/11} \sim 1$ is required. Slippage factors of order $\sim 1$  lead to time periods on which the r-mode is unstable with a timescale of at most 1000 yr, which is about $10^{-3}$ times shorter than the accretion timescale. This would mean that only about 1 in 1000 LMXBs in the galaxy are possible LIGO sources. However, lower slippage factors lead to a longer duration of the gravitational wave emission, but also lower frequencies. We also note that in this model we have considered only very fast accretors  with $\dot{M} \sim 10^{-8} M_\odot \rm{yr}^{-1}$ and most LMXBs in our galaxy accrete at slower rates. Investigations with more accurate nuclear heating models are a subject for future work.

 Our analysis could be made more realistic in several ways, such as
by including the effects of magnetic fields, compressibility,
multi-fluid composition \cite{AnderssonSup}, superfluidity, superconductivity, etc.
These features would render the model more realistic, but its
generic features ought to persist, since the upshot would still be
a dense set of mode frequencies exhibiting three mode resonances
and parametric instabilities with low threshold amplitudes. 
Although the behavior of the star would differ quantitatively
in a model different from ours in detail, 
we expect the qualitative behaviors we have found to be robust, 
as they are well described by quasi-stationary mode evolutions
whose slow variations are determined by competitions between
dissipation and neutrino cooling, and accretion spin-up
and gravitational radiation spin-down. In our model, it seems
that three mode evolution involving interactions of the
r-mode with two daughters at the lowest parametric instability
threshold is often sufficient to quench the instability. Our
treatment is inadequate to follow what happens when the system
runs away; for this, coupling to additional modes is essential.
For this regime, a generalization of the work of Brink {\it et al.} 
\cite{Jeandrew1, Jeandrew2, Jeandrew3} that includes accretion spin-up, viscous heating and
neutrino cooling would be needed. Such a calculation is formidable
even in a ``simple" model involving coupled inertial modes of an
incompressible star.

\section*{Acknowledgments}
It is a pleasure to thank Jeandrew Brink and \'{E}anna Flanagan for useful discussions. RB would especially like to thank Jeandrew for useful discussions, encouragement and advice at the beginning of this project, without which the project would not have been started. RB is very grateful to Gregory Daues for steady encouragement and support for the duration of this project, and also to Gabrielle Allen and Ed Seidel. This research was funded by grants NSF AST-0307273, NSF AST-0606710 and NSF PHY-0354631.
  \renewcommand{\theequation}{A-\arabic{equation}}
  % redefine the command that creates the equation no.
  \setcounter{equation}{0}  % reset counter 
  \section*{APPENDIX A}  % use *-form to suppress numbering
This appendix will sketch the derivation of Eqs.\ (\ref{Amplitudes}) from the Lagrangian density. We follow closely Appendix A in Schenk {\it et al.}, which contains the derivation of the equations of motion for constant $\Omega$.

The Lagrangian density as given by Eq.\ (A1) in Schenk {\it et al.} \cite{Schenk} is 
\begin{equation}
{\cal L} = \frac{1}{2} \dot{{\bf \xi}} \cdot \dot{{\bf \xi}} + \frac{1}{2} \dot{{\bf \xi}} \cdot {\bf B} \cdot {\bf \xi} - \frac{1}{2} {\bf \xi} \cdot {\bf C} \cdot {\bf \xi} + {\bf a}_{\rm{ext}}(t)\cdot {\bf \xi},
\end{equation}
where the operators ${\bf B} \cdot {\bf \xi} = 2 {\bf \Omega} \times {\bf \xi}$ and 
\begin{eqnarray} 
\rho (C \cdot \xi)_i &=& - \nabla_i (\Gamma_1 p \nabla_j \xi^j) + \nabla_i p \nabla_j \xi^j + \rho \nabla_i \delta \phi \;\;\;\;\;\;
\\ \nonumber  &-& \nabla_j p \nabla_i \xi^j + \rho\xi^j \nabla_j \nabla_i \phi + \rho \xi^j \nabla_j \nabla_i \phi_{\rm{rot}}
\end{eqnarray}
 with $\phi_{\rm{rot}}= - (1/2) ({\bf\Omega} \times {\bf x})^2$.
%The action is
%\begin{equation}
%S = \int d^3 r \rho({\bf r}) \int dt {\cal L}
%\end{equation}
 We are interested in a situation where the uniform angular velocity of the star changes slowly on the timescale of the rotation period itself. In order to remove the time dependence we define the new displacement and time variables
\begin{eqnarray}
{\bf \xi} = \frac{\tilde{{\bf \xi}}}{\sqrt{\Omega}}, \;\;\;\; d\tau = \Omega dt .
\end{eqnarray} 
In terms of these new variables the Lagrangian density can be written as
\begin{eqnarray}
\tilde{{\cal L}} &=& \frac{1}{2} {\bf \tilde{\xi}'} \cdot  {\bf \tilde{\xi}'} +  \frac{1}{2} {\bf \tilde{\xi}'} \cdot ( {\bf \tilde{B}} \cdot {\bf \tilde{\xi}}) + \frac{(\sqrt{\Omega})''}{2 \sqrt{\Omega}} |{\bf \tilde{\xi}}|^2  \\ \nonumber
&-& \frac{1}{2} {\bf \tilde{\xi}} \cdot {\bf \tilde{C}} \cdot {\bf \tilde{\xi}} + \frac{{\bf a}_{\rm{ext}}(t)}{\Omega^{3/2}} \cdot {\bf \tilde{\xi}},
\end{eqnarray}
where the primes denote derivatives with respect to $\tau$, ${\bf \tilde{B}} = \Omega^{-1} B$ and ${\bf \tilde{C}} = \Omega^{-2} C$.
The momentum canonically conjugate to ${\bf \tilde{\xi}}$ is 
\begin{equation}
{\bf \tilde{\pi}} = \frac{\partial {\cal L}}{\partial {\bf \tilde{\xi}'}} =  {\bf \tilde{\xi}'} + {\bf \hat{\Omega}} \times {\bf \tilde{\xi}}.
\end{equation}
The associated Hamiltonian density is
\begin{equation}
{\cal H} = \frac{1}{2} \left|{\bf \tilde{\pi}} - \frac{1}{2} {\bf \tilde{B}} \cdot {\bf \tilde{\xi}}\right|^2 -  \frac{(\sqrt{\Omega})''}{2 \sqrt{\Omega}} |{\bf \tilde{\xi}}|^2 + \frac{1}{2}  {\bf \tilde{\xi}} \cdot {\bf \tilde{C}} \cdot {\bf \tilde{\xi}} - \frac{{\bf a}_{\rm{ext}}}{\Omega^{3/2}} \cdot {\bf \tilde{\xi}}.
\end{equation}
Hamilton's equations of motions can be written as
\begin{equation}
\label{motion}
{\bf\tilde{\zeta}'} = T \cdot  {\bf \tilde{\zeta}} + {\bf F}(\tau),
\end{equation}
where
\[ \zeta = \left( \begin{array}{c}
 {\bf\tilde{\xi}} \\
 {\bf \tilde{\pi}}
 \end{array}
 \right),
\]
the operator $T$ is $T = T_0 +T_1$ with
\[
T_0 = \left(\begin{array}{cc}
-\frac{1}{2} {\bf \tilde{B}} & 1 \\
\frac{1}{4} {\bf \tilde{B}}^2 - {\bf \tilde{C}} & -\frac{1}{2} {\bf \tilde{B}}
\end{array}
\right)
\] and
\[T_1 =
 \left(\begin{array}{cc}
0 & 0 \\
 \frac{(\sqrt{\Omega})''}{\sqrt{\Omega}}  & 0
\end{array}
\right),
\]
and
\[
{\bf F}(\tau) = \left(\begin{array}{c}
0\\
\frac{{\bf a}_{\rm{ext}}}{\Omega^{3/2}}
\end{array}
\right).
\]
We assume solutions of the form $\tilde{\zeta}(\tau,{\bf x}) = e^{i \tilde{\omega} t} \tilde{\zeta}({\bf x})$. Specializing to the case of no forcing term $a_{\rm{ext}} = 0$ leads to the eigenvalue equation
\begin{equation}
(T_0 - i \tilde{\omega}) \tilde{\zeta}({\bf x}) = 0.
\end{equation}
Since the operator $T_0$ is not Hermitian it will have distinct right and left eigenvectors. Similar to Schenk {\it et al.} \cite{Schenk} we label the right eigenvectors of T as $\tilde{\zeta}_A$, and the associated eigenfrequencies as $\tilde{\omega}_A = \omega_A/\Omega$, and the eigenvalue equation above becomes
\begin{equation}
\label{eigen}
(T_0 - i \tilde{\omega}_A) \tilde{\zeta}_A({\bf x}) = 0.
\end{equation}
The left eigenvectors $\chi_A$ satisfy
\begin{equation}
(T_0^\dagger - i \tilde{\omega}_A^\star) \tilde{\chi}_A = 0,
\end{equation}
where
\[
T_0^\dagger = \left(\begin{array}{cc}
\frac{1}{2} {\bf \tilde{B}} & \frac{1}{4} {\bf \tilde{B}}^2 - {\bf \tilde{C}} \\
1 &  \frac{1}{2} {\bf \tilde{B}}
\end{array}
\right)
\] 
For simplicity, in this appendix we specialize to the case of no Jordan chains when the set of right eigenvectors forms a complete basis. The orthonormality relation between right and left eigenvectors is 
\begin{equation}
\label{norma}
\left<{\bf \tilde{\chi}}_A,{\bf \tilde{\zeta}}_B\right> = \int d^3{\bf x} \rho({\bf x}) {\bf \tilde{\chi}}_A^\dagger \cdot {\bf \tilde{\zeta}}_B = \delta_{AB}.
\end{equation}
 We can expand $\zeta(\tau, {\bf x})$ in this basis as
\begin{equation}
\label{expansion}
\zeta(\tau,{\bf x}) = \sum_A C_A(\tau) \zeta_A({\bf x}),
\end{equation}
where the coefficients $C_A$ are given by the inverse of this mode expansion
\begin{equation}
C_A(\tau) = \left<{\bf \tilde{\chi}}_A, {\bf\tilde{\zeta}}(\tau,{\bf x}) \right>.
\end{equation}
Using Eqs.\ (\ref{expansion},\ref{eigen},\ref{norma}) in Eq.\ (\ref{motion}) leads to the equations of motion for the mode amplitudes
\begin{eqnarray}
C_A' - i \tilde{\omega}_A C_A &=&g(\tau)\sum_B C_B^\star \left<\tilde{\chi}_A, \left( \begin{array}{c}  0\\
{\bf \tilde{\xi}}_B \end{array}\right) \right> \\ \nonumber
&+&\left<{\bf \tilde{\chi}}_A,F(\tau) \right>,
\end{eqnarray}
where $g(\tau) =(\sqrt{\Omega})''/\sqrt{\Omega}$. Following Sec.\ IV of Schenk {\it et al.} \cite{Schenk} we replace the externally applied acceleration by the nonlinear acceleration given by Eq.\ (4.2) of Ref.\ \cite{Schenk}. The inner product can be written in terms of the displacement variable $\tilde{\xi}$. The left eigenvectors are
\[
{\bf \tilde{\chi}}_A =\left( \begin{array}{c} {\bf \tilde{\sigma}}_A \\
{\bf \tilde{\tau}}_A
\end{array} \right),
\]
 where ${\bf \tilde{\tau}}_A$ can be chosen to be proportional to ${\bf \tilde{\xi}}_A$ because they satisfy the same matrix equation.
  \begin{equation}
  {\bf \tilde{\tau}}_A = - i {\bf \tilde{\xi}}_A/\tilde{b}_A,
  \end{equation}
   which corresponds to Eq.\ (A-45) in Schenk {\it et al.} \cite{Schenk} with the proportionality constant $\tilde{b}_A = \Omega^{-1} b_A = M R^2/\tilde{\omega}_A$ (also given by Eq.\ (2.36) of Ref.\ \cite{Schenk}).

 The equations of  motion for the mode amplitudes become
 \begin{eqnarray}\label{CA2}
C_A' - i \tilde{\omega}_A C_A &=&\frac{i g(\tau)}{\tilde{b}_A} \sum_B C_B  \int d^3 {\bf x} \tilde{\xi}_A^\star \cdot \tilde{\xi}_B   \\ \nonumber
&+& \frac{i M R^2}{\tilde{b}_A} \sum_{BC}  \tilde{\kappa}_{ABC}^\star C_B^\star C_C^\star,
\end{eqnarray}
where the nonlinear coupling $\tilde{\kappa}_{ABC} = \kappa_{ABC}/(M R^2\Omega^2)$ and  $\kappa_{ABC}$ is explicitly give by Eq.\ (4.20) of Ref.\ \cite{Schenk}. The $g(\tau)$ integral mixes only modes with $m_A = m_B$ because of the $e^{i m\phi}$ dependence of the displacement eigenvectors $\tilde{\xi}$. ($\int_0^{2 \pi} \; d\phi  e^{i (m_A - m_B) \phi} = 0$ if  $m_A \ne m_B$.) So, this term will be zero for our mode triplet. Also, in the case of a single mode triplet there is only one coupling and Eqs.\ (\ref{CA2}) take the form of Eqs.\ ({\ref{Amplitudes}).
 %${\bf \tilde \sigma}_A = i \tilde{\omega}_A  {\bf \tilde{\tau}}_A - (1/2) {\bf \tilde{B}} \cdot {\bf \tilde{\tau}}_A$ and.
\renewcommand{\theequation}{B-\arabic{equation}}
  % redefine the command that creates the equation no.
  \setcounter{equation}{0}  % reset counter 
  \section*{APPENDIX B}  % use *-form to suppress numbering
%\subsection*{Behavior Near Thermal Equilibrium}
In this appendix we study the behavior of the mode amplitudes and temperature near equilibrium assuming constant angular velocity.  We are performing a first order expansion of Eqs.\ (\ref{phase}) and (\ref{thermalevol}).   Similar to Ref.\ \cite{WHL}, each of the five variables is expanded about its equilibrium $(X_j)_e$ as follows
\begin{equation}
X_j(\tilde{\tau}) = \{|\bar{C}_\alpha|, |\bar{C}_\beta|, |\bar{C}_\gamma|, \phi, T_8\} = (X_j)_e [1 + \zeta_j(\tilde{\tau})]
\end{equation}
where the perturbation $|\zeta_j| << 1$ and $j = \alpha, \beta, \gamma, T$. 
%The full expansion gives 
%\begin{equation}
%\frac{d \zeta^i}{d \tilde{\tau}} = A^{i j} \zeta^j
%\end{equation}
The expansion leads to a first order differential equation for each $\zeta_j$
\begin{eqnarray}
\label{expansion}
\frac{d\zeta_\alpha}{d \tilde{\tau}} &=& \frac{(\tilde{\gamma}_\alpha)_e}{\tilde{\Omega} |\delta \tilde \omega|}  \left[\zeta_\alpha -\zeta_\beta -\zeta_\gamma - \left(\frac{\phi}{\tan \phi}\right)_e \zeta_\phi \right. \\ \nonumber
&-& \left. \left(\frac{T_8}{\tilde{\gamma}_\alpha}\right)_e \left(\frac{\partial \tilde{\gamma}_\alpha}{\partial T_8}\right)_e\zeta_T\right], \\ \nonumber
\frac{d\zeta_\beta}{d \tilde{\tau}} &=& \frac{(\tilde{\gamma}_\beta)_e}{\tilde{\Omega} |\delta \tilde \omega|}  \left[\zeta_\alpha - \zeta_\beta + \zeta_\gamma + \left(\frac{\phi}{\tan \phi}\right)_e \zeta_\phi \right. \\ \nonumber
&-& \left. \left(\frac{T_8}{\tilde{\gamma}_\beta}\right)_e \left(\frac{\partial \tilde{\gamma}_\beta}{\partial T_8}\right)_e\zeta_T\right], \\ \nonumber
\frac{d\zeta_\gamma}{d \tilde{\tau}} &=& \frac{(\tilde{\gamma}_\gamma)_e}{\tilde{\Omega} |\delta \tilde \omega|}  \left[\zeta_\alpha +\zeta_\beta -\zeta_\gamma + \left(\frac{\phi}{\tan \phi}\right)_e \zeta_\phi \right. \\ \nonumber
&-& \left. \left(\frac{T_8}{\tilde{\gamma}_\gamma}\right)_e \left(\frac{\partial \tilde{\gamma}_\gamma}{\partial T_8}\right)_e\zeta_T\right], \\ \nonumber
\frac{d\zeta_\phi}{d \tilde{\tau}} &=& \frac{1}{\phi_e \tan \phi_e} \left(\zeta_\alpha \frac{\tilde{\gamma}_\alpha+\tilde{\gamma}_\beta+\tilde{\gamma}_\gamma}{\tilde{\Omega} |\delta \tilde \omega|} + \zeta_\beta \frac{- \tilde{\gamma}_\alpha-\tilde{\gamma}_\beta+\tilde{\gamma}_\gamma}{\tilde{\Omega} |\delta \tilde \omega|}\right. \\ \nonumber 
&+& \left. \zeta_\gamma \frac{- \tilde{\gamma}_\alpha+\tilde{\gamma}_\beta-\tilde{\gamma}_\gamma}{\tilde{\Omega} |\delta \tilde \omega|}\right)_e +  \frac{(\tilde{\gamma}_\alpha-\tilde{\gamma}_\beta-\tilde{\gamma}_\gamma)_e}{\tilde{\Omega} |\delta \tilde \omega|} \zeta_\phi, \\ \nonumber
\frac{d\zeta_T}{d \tilde{\tau}} &=& \frac{M R^2\Omega_c^2 \tilde\gamma_\alpha \tilde\gamma_\beta \tilde \gamma_\gamma}{2 \tilde \kappa^2 \tilde \omega_\alpha \tilde \omega_\beta \tilde \omega_\gamma \tilde \Omega |\delta \tilde \omega| C(T_e) T_{8e}} \left(1 + \frac{1}{\tan \phi_e^2}\right) \\ \nonumber
&\times&\left[2\left(\tilde \omega_\alpha \frac{\tilde \gamma_{\alpha\, v}}{\tilde \gamma_\alpha} \zeta_\alpha +  \tilde\omega_\beta \zeta_\beta + \tilde\omega_\gamma \zeta_\gamma \right) \right. \\ \nonumber
&+& \left. T_{8e} \left(\tilde\omega_\alpha\frac{1}{\tilde\gamma_\alpha} \frac{\partial \tilde \gamma_\alpha}{\partial T_8} + \tilde\omega_\beta \frac{1}{\tilde\gamma_\beta} \frac{\partial \tilde \gamma_\beta}{\partial T_8} + \tilde \omega_\gamma \frac{1}{\tilde\gamma_\gamma} \frac{\partial \tilde \gamma_\gamma}{\partial T_8} \right)_e \zeta_T \right] \\ \nonumber
&-& \left(\frac{d L_\nu}{dT_8}\right)_e \frac{1}{\Omega_c \tilde\Omega |\delta \tilde \omega| C(T_e)}\zeta_T,
\end{eqnarray}
where the equilibrium amplitudes $|C_j|_e$ have been written in terms of the corresponding driving and damping  rates  using Eqs.\ (\ref{stationarySol}). Eq.\ (\ref{expansion}) can be written in matrix form as
\begin{equation}
\frac{d \zeta_j}{d \tilde{\tau}} = A^{i j} \zeta_i.
\end{equation}
Let $\zeta_j \propto \exp(\lambda \tilde{\tau})$. The determinant $||A^{i j} - \lambda \delta^{i j}|| = 0 $ leads to the eigenvalue equation
\begin{equation}
\lambda^5 + a_4 \lambda^4 + a_3 \lambda^3 + a_2 \lambda^2 + a_1 \lambda + a_0  = 0.
\end{equation}
The coefficients $a_j$ with $j = 0,4$ are
\begin{eqnarray}
a_4 &=& 2 \tan \phi_e = \frac{\tilde \gamma_\beta + \tilde \gamma_\gamma - \tilde \gamma_\alpha}{\tilde \Omega |\delta \tilde \omega|}, \\ \nonumber
a_3 &\approx& \frac{2}{\tan \phi_e^2} \frac{\tilde \gamma_\beta^2 + \tilde \gamma_\gamma^2 + \tilde \gamma_\alpha^2}{(\tilde \Omega |\delta \tilde \omega|)^2} + \tan \phi_e^2 - 1, \\ \nonumber
a_2 &\approx& \frac{\tilde \gamma_\alpha \tilde \gamma_\beta \tilde \gamma_\gamma}{(\tilde \Omega |\delta \tilde \omega|)^3} \left(\frac{12}{\tan \phi_e^2}+1\right), \\ \nonumber
a_1 &\approx& \frac{4 \tilde \gamma_\alpha \tilde \gamma_\beta \tilde \gamma_\gamma}{(\tilde \Omega |\delta \tilde \omega|)^3} \left(\frac{1}{\tan \phi_e}+\tan \phi \right), \\ \nonumber
a_0 &\approx& \frac{2 M R^2 \Omega_c^2}{\tilde \kappa^2 \tilde \omega_\alpha \tilde \omega_\beta \tilde\omega_\gamma C(T_e)} \frac{(\tilde \gamma_\alpha \tilde \gamma_\beta \tilde \gamma_\gamma)^2}{(\tilde \Omega |\delta \tilde \omega|)^4} \frac{1}{\tan\phi_e} \left(1 + \frac{1}{\tan\phi_e^2}\right) \\ \nonumber
&\times&\left[\frac{\tilde \omega_\alpha}{\tilde \gamma_\alpha} \left(\frac{\partial \tilde \gamma_\alpha}{\partial T_8}\right)_e+\frac{\tilde \omega_\beta}{\tilde \gamma_\beta} \left(\frac{\partial \tilde \gamma_\beta}{\partial T_8}\right)_e+\frac{\tilde \omega_\gamma}{\tilde \gamma_\gamma} \left(\frac{\partial \tilde \gamma_\gamma}{\partial T_8}\right)_e \right]  \\ \nonumber
&-&\frac{4 \tilde \gamma_\alpha \tilde \gamma_\beta \tilde \gamma_\gamma}{(\tilde \Omega |\delta \tilde \omega|)^3} \frac{1}{\tan\phi_e} \frac{1}{\tilde \Omega |\delta \tilde \omega|C(T_e)} \left(\frac{d L_\nu}{d T_8}\right)_e.
\end{eqnarray}
%The coefficients $a_0 = O(\gamma^3)$ and $a_1 = O(\gamma^2)$ are smaller than $a_{2,4} = O(\gamma)$ and $a_3 = O(1)$. 
The eigenvalues can be approximated as
\begin{eqnarray}
\lambda_{1,2} &\approx& - \frac{a_4}{2} - \epsilon \pm i \sqrt{\frac{a_1}{\epsilon^2 + w^2} - \left(\frac{a_4}{2} + \epsilon\right)^2}, \\ \nonumber
\lambda_{3,4} &\approx& \epsilon \pm i w, \\ \nonumber
\lambda_5 &\approx& - \frac{a_0}{a_1}, 
\end{eqnarray}
where $\epsilon = (a_2 -a_3 a_4)/a_4$ and $w =  \sqrt{a_1/a_3}$.
The system is unstable when $a_2 - a_3 a_4 > 0$ or $a_0 <0$. The first two eigenvalues will have a negative real part as long as $\tilde \gamma_\beta + \tilde \gamma_\gamma > \tilde \gamma_\alpha$. 
If the heating compensates the cooling of the star $a_0 \approx 0$ and becomes negative if the star can not reach thermal equilibrium. The other critical stability condition $a_2 - a_3 a_4 = 0$ can be written as
\begin{equation}
\left(\frac{\tilde \gamma_\alpha}{\tilde \Omega |\delta \tilde \omega|}\right)^3[1 +  \Gamma_\beta + \Gamma_\gamma -  (\Gamma_\beta^2 + \Gamma_\gamma^2) - (\Gamma_\beta - \Gamma_\gamma)^2 (\Gamma_\beta + \Gamma_\gamma)] =0,
\end{equation}
where $\Gamma_\beta = \gamma_\beta/\gamma_\alpha$ and $\Gamma_\gamma = \gamma_\gamma/\gamma_\alpha$. Note that we have ignored the smaller terms of order $O ([\tilde \gamma_\alpha/(\tilde \Omega |\delta \tilde \omega|)]^5)$. This condition can be rewritten by defining variables $D_1 = \Gamma_\beta + \Gamma_\gamma$ and $D_2 = \Gamma_\beta - \Gamma_\gamma$ 
\begin{equation}
2 + 2 D_1 - D_1^2 - D_2^2 - 2 D_2^2 D_1 = 0.
\end{equation}
If $D_2 =0$ then the equation has one solution $D_1 = 1 + \sqrt{3}$ for $D_1>2$, which corresponds to $\Gamma = \Gamma_\beta = \Gamma_\gamma=1.37$ and matches the result of Wersinger {\it et al.} \cite{Wer}. For the viscosity we consider (see Sec.\ \ref{dissipation}) $a_2 - a_3 a_4 < 0$.

\end{document}